\documentclass[twocolumn,preprintnumbers,amsmath,amssymb,prb,floatfix]{revtex4} 
\usepackage{amssymb,amsmath}
\usepackage{color}
\usepackage{graphicx}
\usepackage{dcolumn}
\usepackage{bm}
\usepackage{upgreek}
\usepackage{marvosym }
\usepackage{latexsym,epsfig}
\usepackage{hyperref}

\begin{document}   

\title{Electronic structure and optical properties of Sr$_2$IrO$_4$ under epitaxial strain}

\author{Churna Bhandari, Zoran S. Popovi\'c and S. Satpathy}

\address{Department of Physics \& Astronomy, University of Missouri,
Columbia, MO 65211, United States of America}
\vspace{10pt}


\begin{abstract}  
We study the modification of the electronic structure in the strong spin-orbit coupled   Sr$_2$IrO$_4$ by epitaxial strain  
using density functional methods. 
Structural optimization shows that strain changes the  internal structural parameters such as the Ir-O-Ir bond angle, which 
has an important effect on  the band structure.
An interesting prediction is the $\Gamma - $X crossover of the valence band maximum with strain, 
while the conduction minimum  at M remains unchanged.
This in turn suggests strong   strain dependence of the transport properties for the hole doped system,
but not when the system is electron doped.  
Taking the measured value of the $\Gamma-X$ separation for the unstrained case, we predict the 
$\Gamma - $X crossover of the valence band maximum to occur for the tensile epitaxial strain  $e_{xx} \approx 3\%$.
A minimal tight-binding model within the $J_{\rm eff} = 1/2$ subspace is developed to describe the main features of the band structure.
The optical absorption spectra under epitaxial strain are computed using density-functional theory, which explains the  observed anisotropy in the optical spectra with the polarization of the incident light.
We show that the optical transitions between the Ir (d) states, which are dipole forbidden, can be explained in terms of the admixture 
of Ir (p) orbitals with  the Ir (d) bands.


\end{abstract}

%
%
%
\maketitle
%
%

\section{Introduction}

The   5d oxides such as Sr$_2$IrO$_4$ (SIO) are of considerable current interest due to the presence of a large spin-orbit coupling (SOC)  which leads to many  novel features such as the  spin-orbit assisted $J_{\rm eff} = 1/2  $ Mott insulator\cite{KimPRL08} and spin-orbital entangled electron states.
It has been suggested that the spin-orbital entanglement induced by the strong SOC in these structures could 
make these materials hosts for several unconventional features such as  the Kitaev model\cite{JackeliPRL09,Balentsnat10}, 
quantum spin Hall effect at room temperature,\cite{ShitadePRL09} or unconventional superconductivity\cite{WangPRL011,WantanbePRL13}. 

Strain is an important parameter for probing the nature of the  electron states, and it can be induced by pressure or by epitaxial growth on  lattice-matched substrates. 
Tuning of the band gap is an important aspect for functional manipulation for potential device applications.  

There have been several studies of SIO under epitaxial strain condition, both from theory and experiments.
Samples of SIO have been grown epitaxially on a number of substrates such as SrTiO$_3$, LaAlO$_3$, GdScO$_3$, etc.\cite{RyanPRB13, NicholsAPL13, Lupascu}.
Resistivity and optical absorption measurements on these structures have shown that the  Mott-Hubbard gap is preserved under epitaxial strain, but its magnitude can be tuned by varying the strain. The changes in the electronic structure show up in the optical properties
as red or blue shift of the optical absorption under strain condition\cite{NicholsAPL13, Nichols2-APL13}.  
Several  theoretical works have also addressed the electronic structure of SIO under strain\cite{ZhangPRL13, LadoPRB015, Kimsr016}. 
The optical properties  were studied by 
Zhang et al.\cite{ZhangPRL13} and Kim et al.\cite{Kimsr016}. 
These calculations were limited to the low energy range (0 - 2 eV) and, furthermore, the polarization dependence
 of the optical absorption
has not been studied theoretically,
even though experiments show a strong anisotropy for absorption with  E $ \parallel {\rm plane}$ vs. E $ \parallel \hat z$ .
Thus, a full understanding of the electronic and optical properties is still missing.

In this paper, we study the effect of the epitaxial strain on the electronic structure and optical absorption of SIO from density-functional theory. We find an interesting $\Gamma-X$ crossover of the valence band top under strain, which we explain from a tight-binding model,
 and also find the anisotropy in the absorption spectrum for light polarized along the plane vs. normal to the plane, in agreement with the experiments. The dipole-forbidden d-d transitions are explained in terms of admixture of the Ir p orbitals with the d bands.

\begin{figure} 
\centering
\includegraphics[scale=0.5]{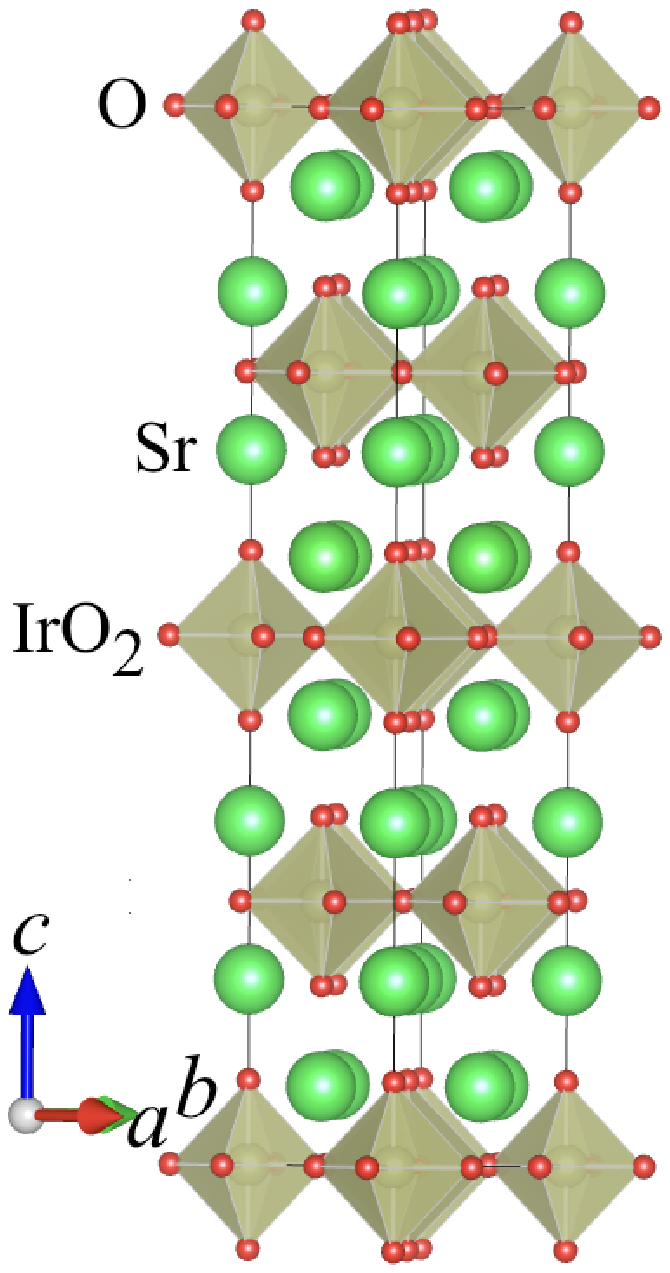}\includegraphics[scale=0.35]{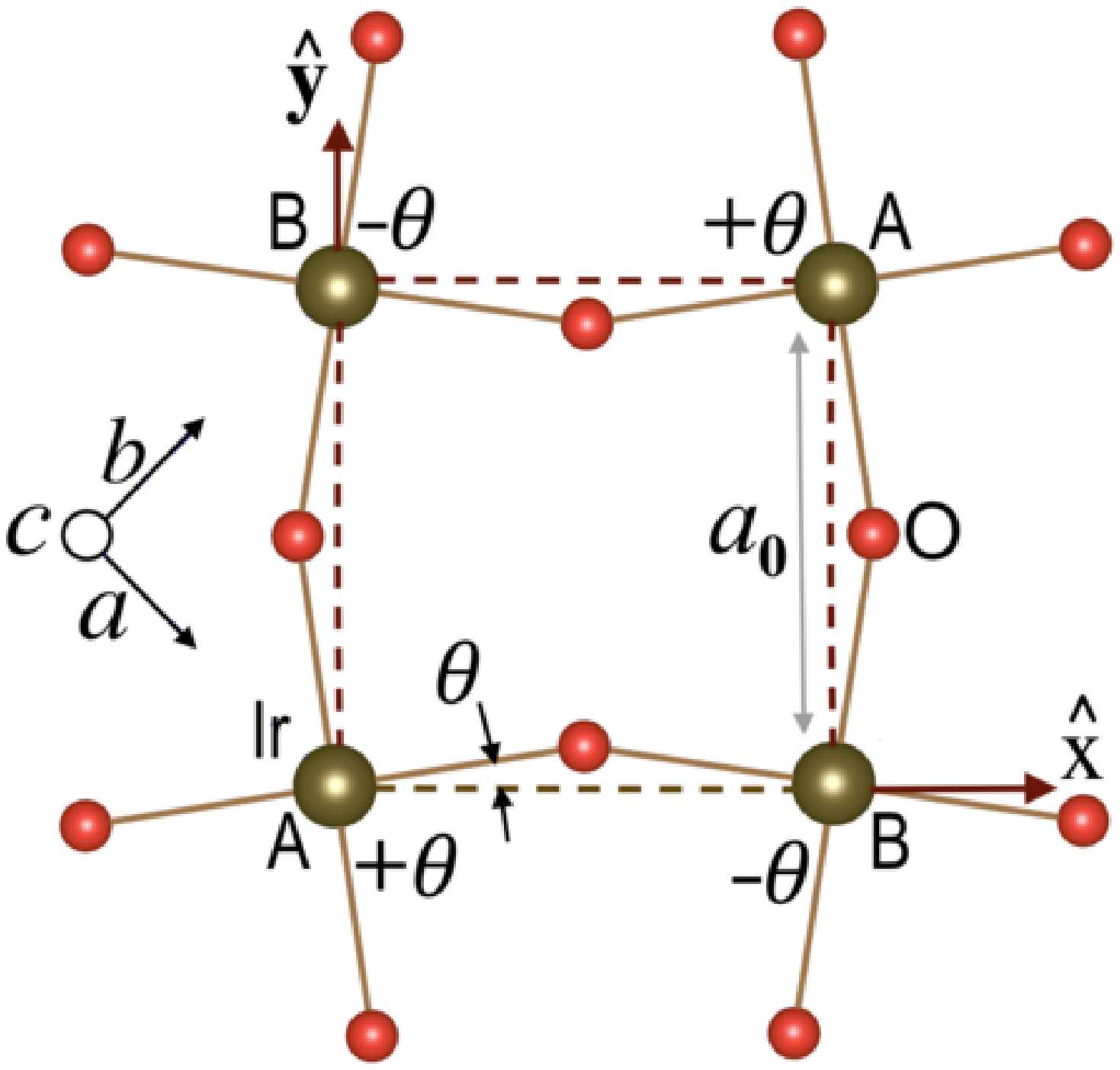}

\caption{ Crystal structure of Sr$_2$IrO$_4$ (left) 
and the IrO$_2$ atom positions on the $ab$-plane (right) indicating the staggered octahedral rotations on the two sublattices. 
Along the $c$ axis, the rotations follow a certain pattern\cite{Crawford}.
}
\label{rotation}
\end{figure}

The organization of the paper is as follows. In Section \ref{S2},
we describe the crystal structure, the method of calculation, and the results of the density-functional structural optimization 
of the crystal structure under strain.
Density-functional results for the band structure and magnetic moments under strain are discussed in Section \ref{dftband}. 
Section \ref{opticsetc} discusses the optical absorption spectrum under strain, and  
the results are  summarized in Section \ref{con}.

\section{Crystal Structure and Method of Calculation} \label{S2}

{\it Bulk Crystal Structure} --
The crystal structure of SIO with space group (142) $I4_1/acd$ consists of eight formula units in the $\sqrt 2 a_0 \times c$ unit cell\cite{Crawford} as shown in Fig. \ref{rotation}, with the lattice constants $a = b = 5.497$ \AA~ and $c = 25.798$ \AA. 
(An equivalent unit cell with the body-centered tetragonal lattice and four formula units in the basis may also be used.) 
Note that $a = \sqrt 2 a_0$.
 The structure shows a staggered rotation of the IrO$_6$ octahedra about the $c$-axis by the angle $\theta = 11.5^{\circ}$. The unit cell has four IrO$_6$ planes stacked along the $c$-axis, each plane consisting of two octahedra,
with staggered rotation angles and antiferromagnetic Ir moments. 
The electronic structure is essentially controlled by the Ir atoms with the $5d^5$ configuration placed in the crystal field of the 
oxygen octahedra, with the Sr atoms playing a passive role of donating electrons to the system, described by the nominal chemical formula Sr$_2^{ 2-}$  Ir$^{4+}$ O$_4^{ 2-}$.
The structure can be thought of as IrO$_2$ layers separated from each other by two intervening SrO planes, making the structure quasi-two dimensional (2D). This means that the basic electronic structure can be modeled by a single plane of the Ir atoms in a minimal model, which we discuss later.

\begin{figure} [hbt]
\centering
\includegraphics[scale=0.5]{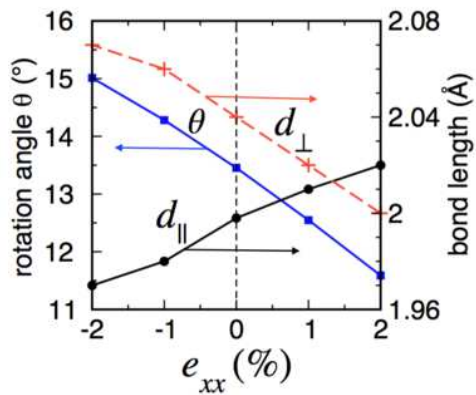}
\caption{ Calculated variation of the octahedral rotation angle  $\theta$ and the in-plane and out-of-plane Ir-O  bond lengths, d$_{\parallel}$ and d$_{\perp}$, as a function of strain. }
\label{theta-strain}
\end{figure}

\begin{table}
\centering
\caption{\label{table1}Structural parameters under strain: In-plane and out-of-plane Ir-O bond lengths, d$_\parallel$ and d$_\perp$, and the octahedral rotation angle $\theta$ calculated from structural relaxation using the density-functional  FP-LMTO method. Lengths are in units of \AA.}
\footnotesize
\begin{tabular}{@{}cccccc}
\hline
$e_{xx}$&+2\% &+1\% & 0 & -1\% & -2\%\\
\hline
a  & 5.61 & 5.55 &  5.497 & 5.44 & 5.39\\
 c    & 24.81 & 25.29 & 25.798 & 26.41 & 26.83\\
 \hline
d$_\parallel$ & 2.02 &2.01 & 2.00 & 1.98 & 1.97 \\  
d$_\perp$ &2.00 &2.02 &2.04 &2.06 &2.07\\                 
$\theta$ (deg.)  & 11.6 & 12.6 & 13.5 & 14.3 & 15.1\\
\hline
\end{tabular}
\end{table}

{\it Density-Functional Methods} --
Density functional theory (DFT)  with linearized full potential muffin-tin orbital method (FP-LMTO)\cite{Methfessel, Kotani10, lmsuite} was used to solve the Kohn-Sham equations within the local spin-density approximation for the exchange and correlational
 functional\cite{vonBarthHedin, Kohn-Sham}.
  The LMTO basis set consisted of spdf orbitals for Ir and Sr and spd orbitals for O, 
which were augmented outside the muffin-tin spheres by two Hankel functions of two different decay lengths.
 The semicore states Ir $5p$ and Sr $3p$ were  treated as valence electrons.  
 The band calculations were carried out for the optimized structures within the local spin density approximation including both the SOC and the Hubbard U terms  (LSDA + SO + U). Following earlier authors, we used $U = 2.7$ eV. We also employed Vienna {\it ab initio} simulation package (VASP) in projector augmented wave (PAW) formalism including Hubbard $U=2$ eV and spin-orbit interaction\cite{KressePRB99} for computing magnetic moments.

{ \it  Structural optimization} --
Atomic positions were relaxed in all our calculations within the local spin-density approximation and the force convergence was obtained with a tolerance of 10$^{-3}$ Ryd/Bohr. For the bulk, unstrained structure, we took the experimental unit cell and optimized
the positions of the atoms. An important structural parameter is the IrO$_6$ octahedral rotation angle $\theta$, which we found to be 
$13.5 ^\circ$ as compared to the measured angle of $11.5 ^\circ$. 

For the epitaxially strained structure, the in-plane lattice constant was changed according to the strain parameter $e_{xx} \equiv (a - a_{\rm bulk})/ a_{\rm bulk}$, while the out-of-plane lattice constant
was determined from the condition that the cell volume is preserved under strain, so that $e_{zz} \equiv (c -c_{\rm bulk})/ c_{\rm bulk} = - 2 e_{xx}$. Structural parameters were then optimized for each strain condition, with the fixed lattice constants. We varied the in-plane strain $e_{xx}$ by $\pm 2 \%$, which is the same order of magnitude as the strains present in the experimental structures, e. g., $e_{xx} \approx 2\%$ for SIO grown on the GdScO$_3$  substrate
and $-2 \%$ for the NdGaO$_3$ substrate\cite{NicholsAPL13}. The computed structural parameters are shown in Table \ref{table1} and
Fig. \ref{theta-strain}. The trend in the structural parameters shown here is similar to the results obtained from an earlier work
using the dynamical mean-field theory\cite{ZhangPRL13}.

For test purposes, we also performed a second set of calculations, where we took the in-plane lattice constant to be the 
same as the substrate as usual, but used the measured c/a ratio from the experiments\cite{NicholsAPL13}.
No significant changes in the bond angles and bond lengths were found.  

\section{Density functional band structure}  \label{dftband}

\subsection{Electronic structure and $\Gamma-X$ crossover}      \label{section-bandstructure}

\begin{figure*} 
\centering
\includegraphics[scale=0.42]{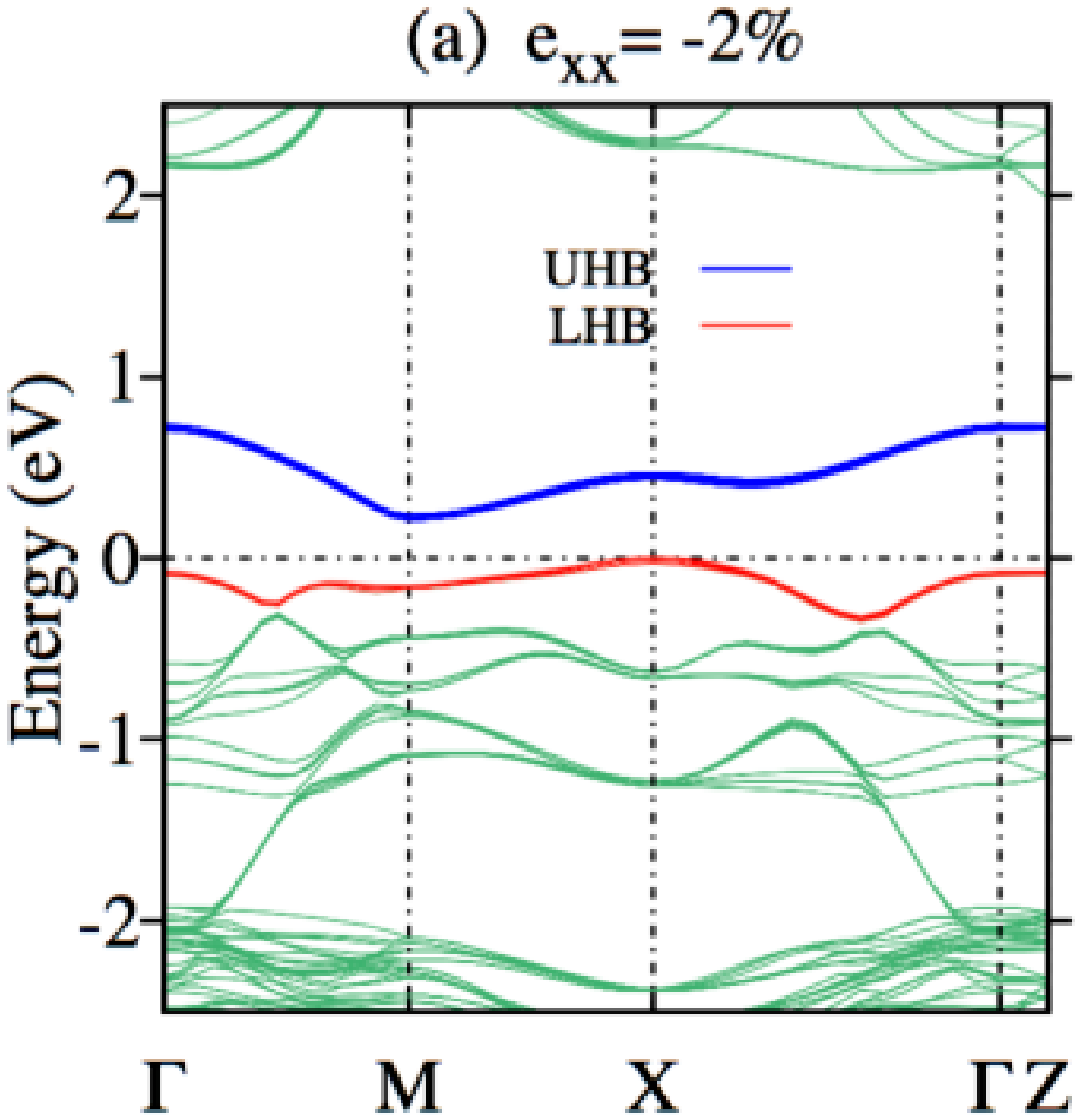} \includegraphics[scale=0.42]{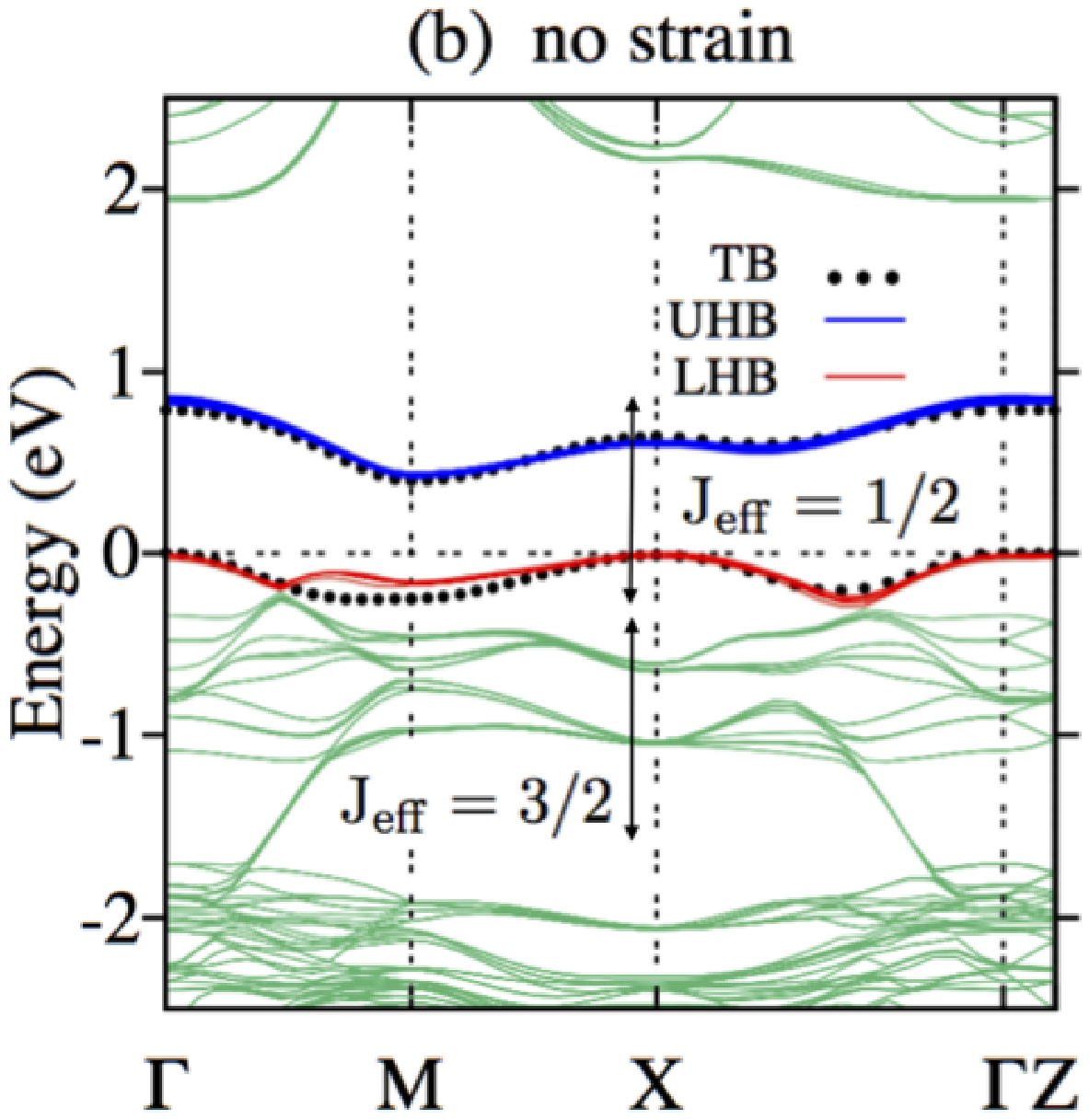}\includegraphics[scale=0.42]{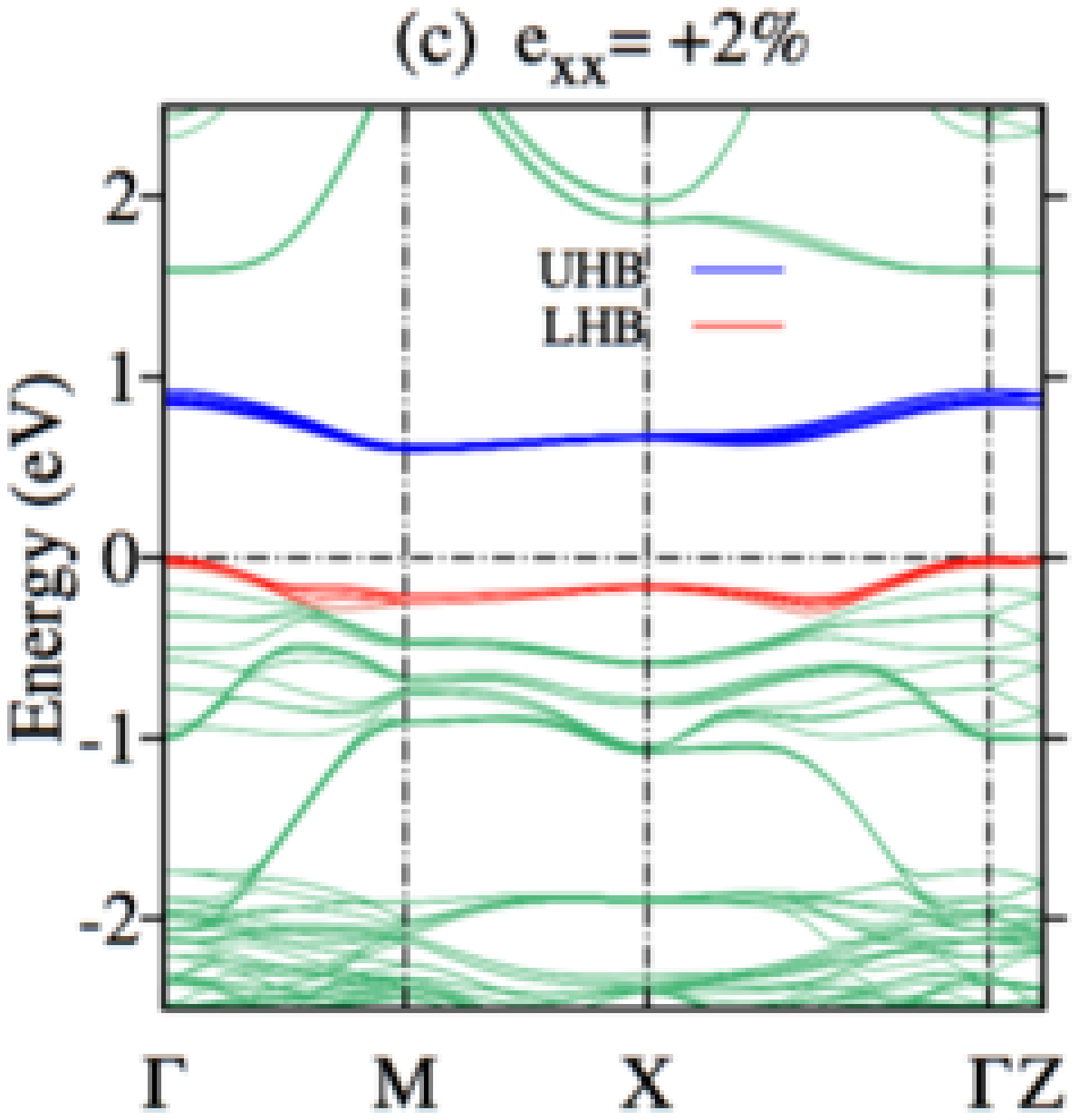}
\caption{ DFT band structures of Sr$_2$IrO$_4$ under strain: a) $e_{xx}$=$-2\%$ (compressive), b) no strain, and c) $e_{xx}$=$+2\%$ (tensile). High symmetry {{\bf k}}-points in the Brillouin-zone are defined as 
$\Gamma$ = (0,0,0), X = $\pi/a_0(1,0,0)$ and M=$\pi/(2a_0) (1,1,0)$ 
and Z = $\pi/c (0,0,1)$, $a_0$ is the Ir-Ir distance on the $ab$-plane. The lattice translation vectors are $\vec{T_1}$ = $a_0 (1,1,0)$, $\vec{T_2}$=$a_0 (-1,1,0)$ and $\vec{T_3}$ = c(0,0,1) with the coordinate system, $\hat{\rm x}$, $\hat{\rm y}$, and $\hat{\rm z}$, chosen along the cube axes. The dotted lines in the middle panel are the fits using the $J_{\rm eff} = 1/2$ tight-binding expression Eq. (\ref{TB-Energy}).} 
\label{band}
\end{figure*}

\begin{figure}
\includegraphics[scale=0.35]{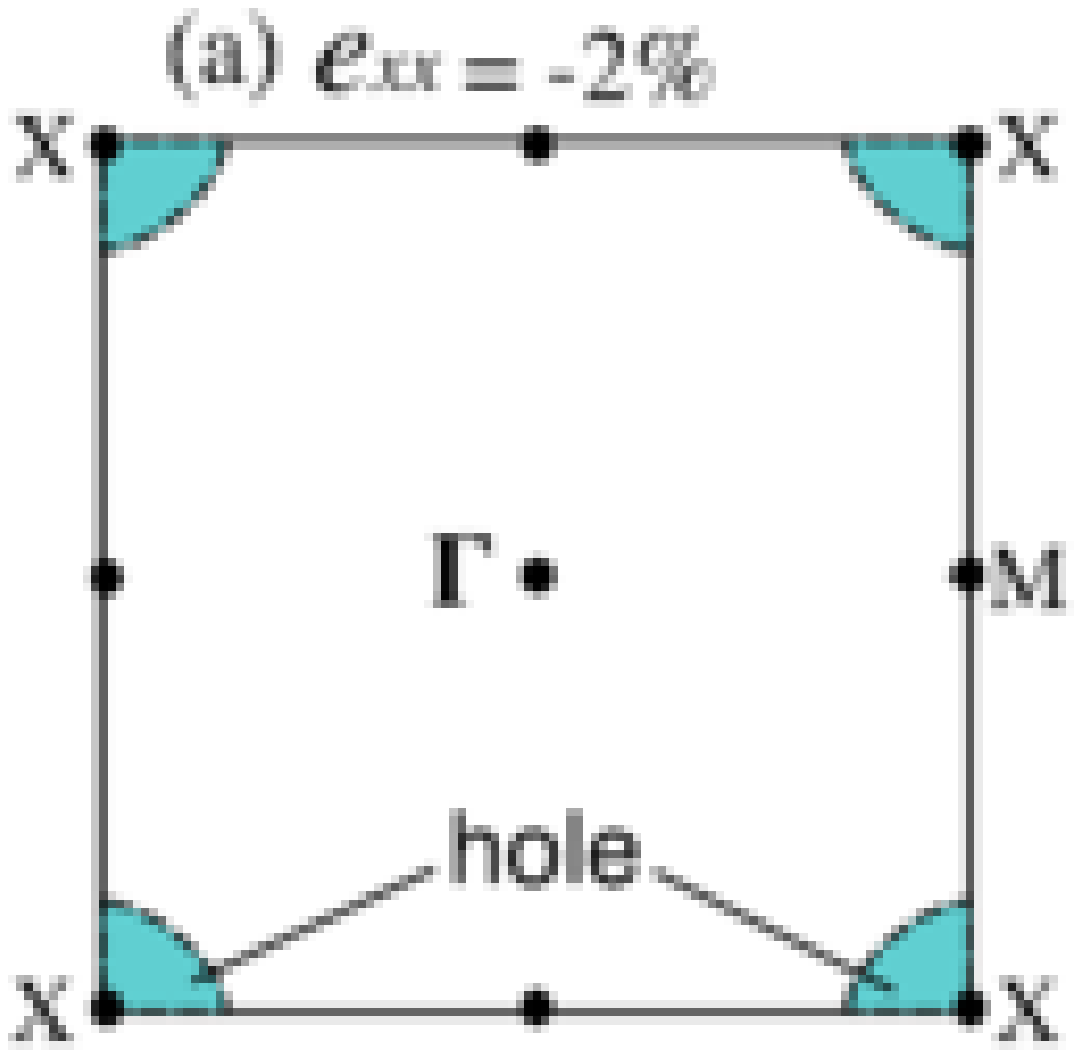}\includegraphics[scale=0.35]{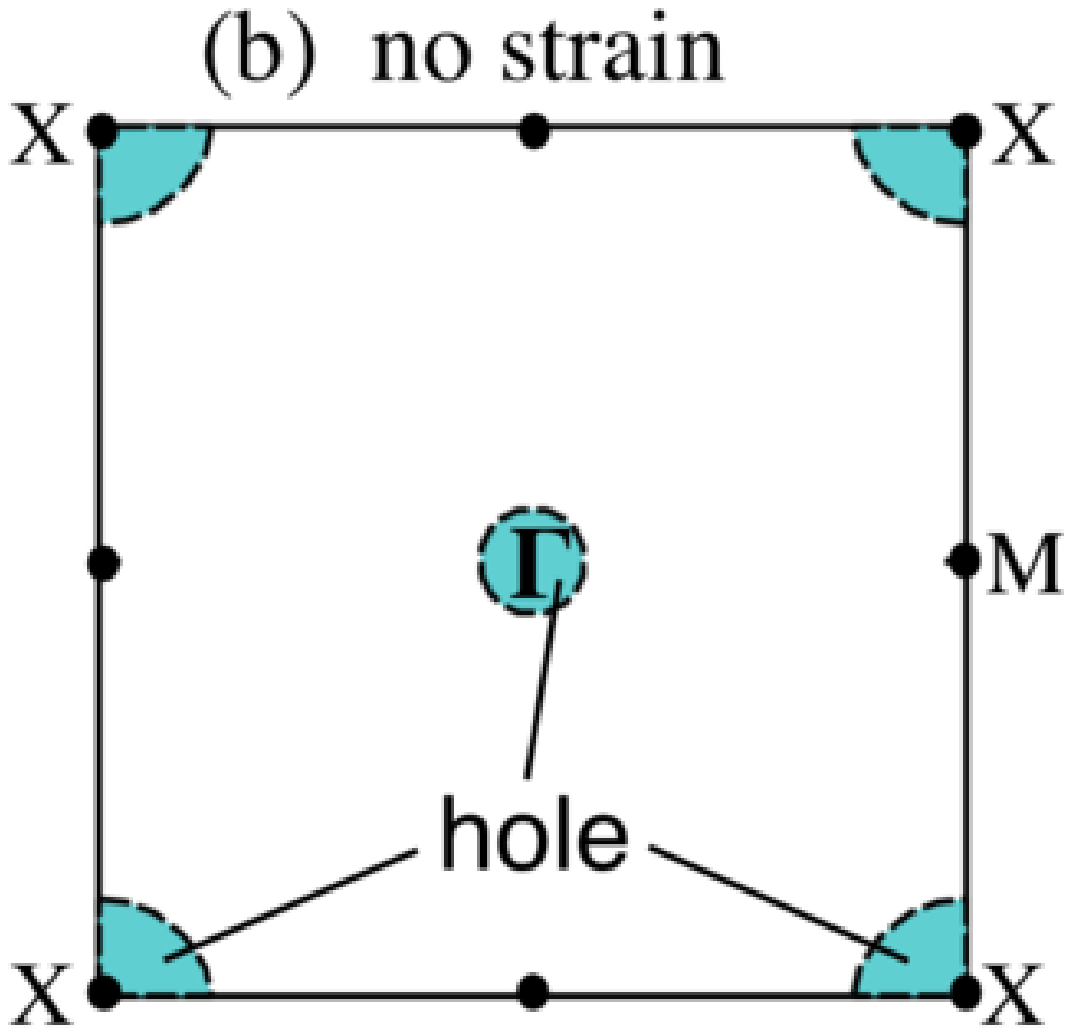}
\includegraphics[scale=0.35]{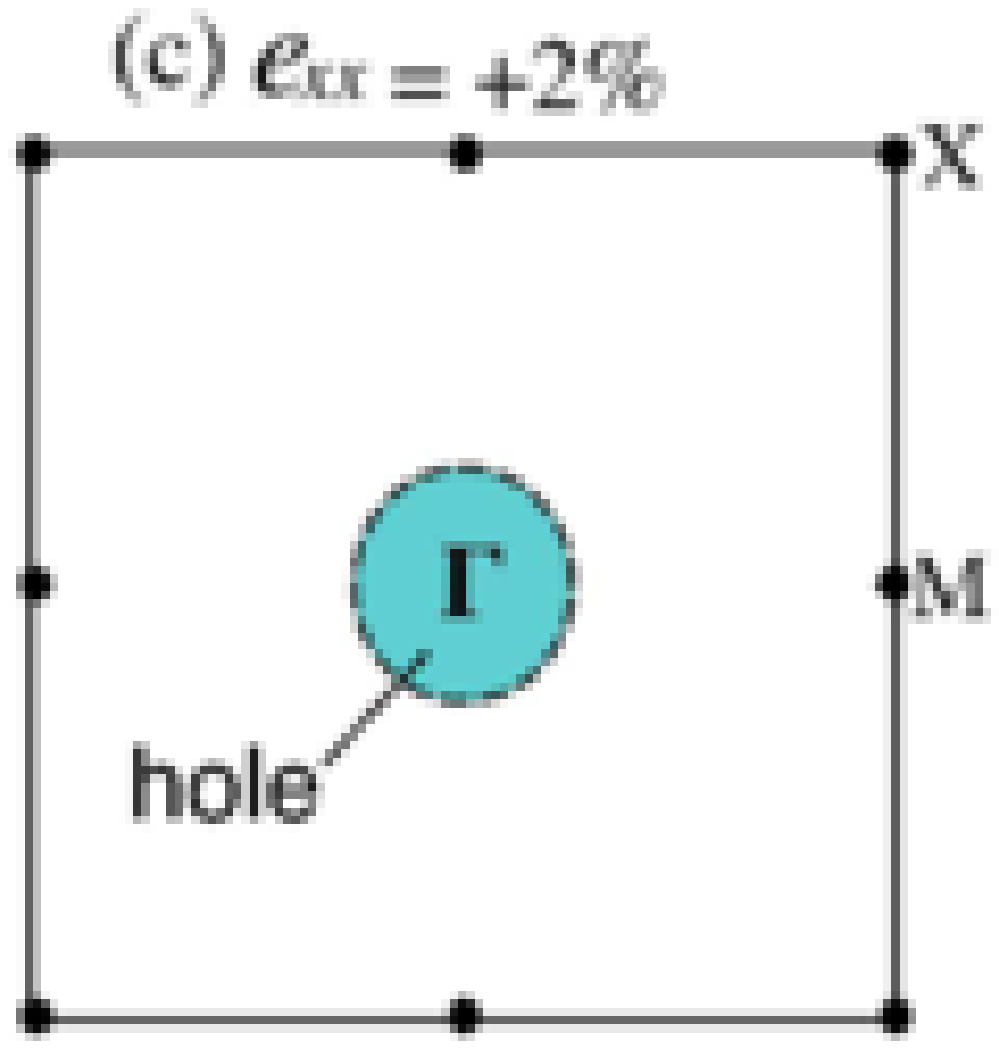}\includegraphics[scale=0.22]{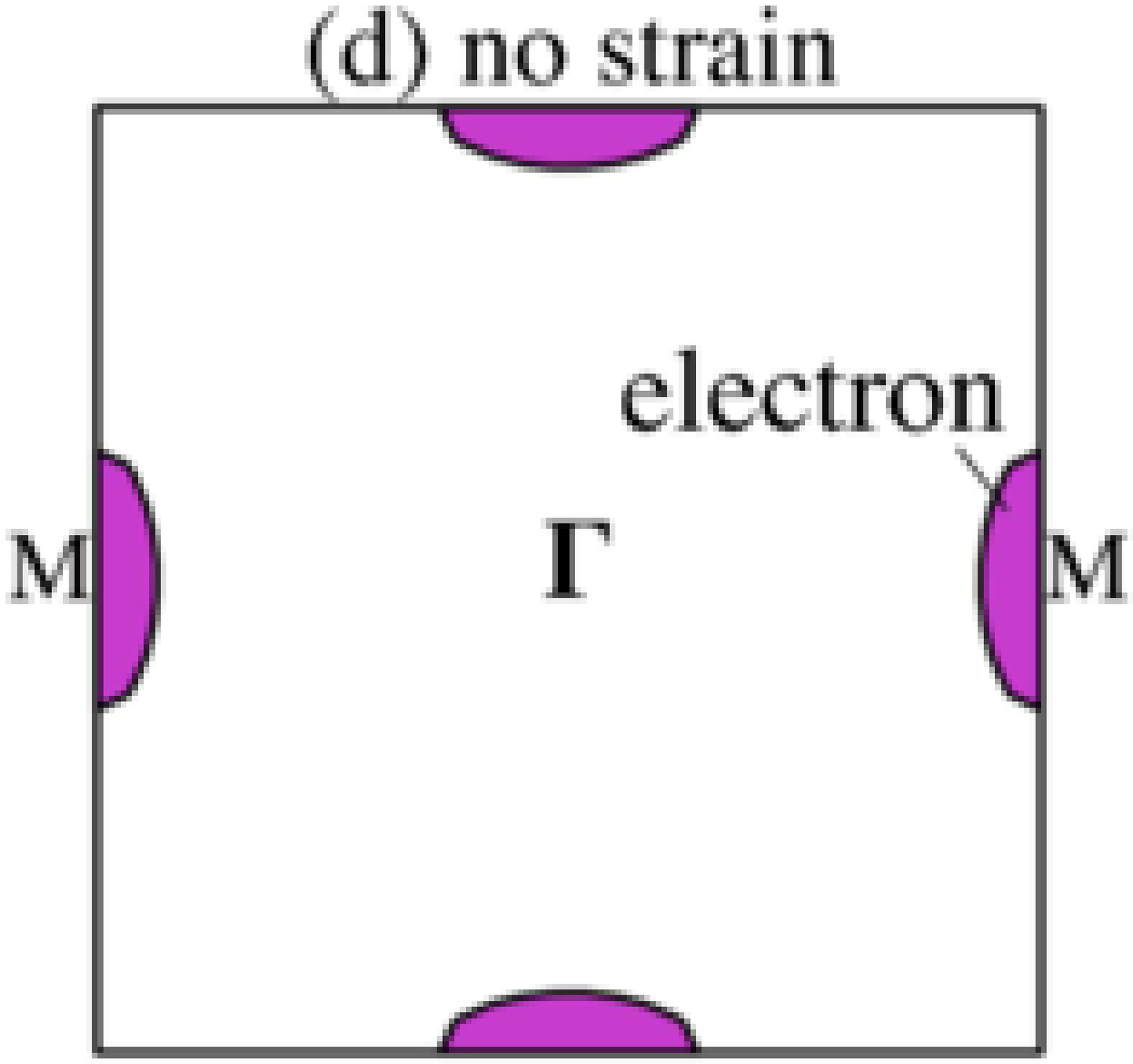}
\caption{ (a) - (c): Fermi surface of hole-doped SIO under epitaxial strain (5\% hole concentration). For compressive strain, 
the hole pocket is at the $X$ point, while for tensile strain, the pocket shifts to $\Gamma$. For the unstrained case, 
valence top at $\Gamma$ is only slightly below $X$, as indicated from the size of the two hole pockets in (b). 
The Fermi surface with 5\% electron doping for the unstrained structure is shown in (d). It remains more or less unchanged with strain,
unlike the hole case, with the elliptical electron pocket occurring at $M$.
}
\label{fermi-h}
\end{figure}

\begin{figure}
\vspace{4cm}\includegraphics[scale=0.55]{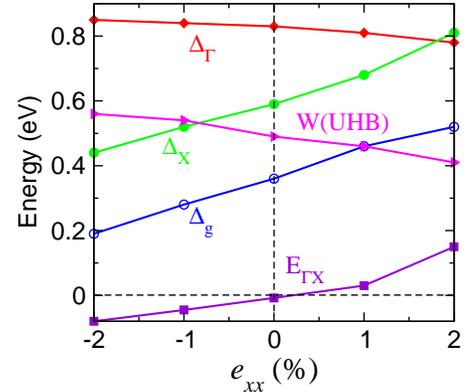}
\caption{Variation of the band gaps with strain $e_{xx}$. Here $\Delta_g$ is the fundamental gap, $\Delta_{\Gamma}$ ($\Delta_{\rm X}$) is the gap at the 
$\Gamma$ ($ {\rm X}$) point, $\rm W $ is the band width of the upper Hubbard band, and
$E_{\Gamma X} = E_\Gamma - E_X$ is the relative energy of the valence band
maximum at $\Gamma$ with respect to the same at $X$.}
\label{gap}
\end{figure}

\begin{figure}
\centering
\includegraphics[scale=0.4]{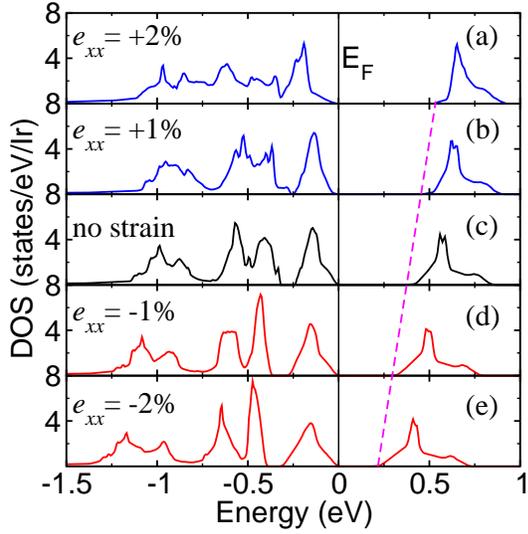}
\caption{ Ir (5d) partial density-of-states as a function of strain near the gap region, indicating the
systematic shift of the band gap.
}
\label{pdos-strain}
\end{figure}

The electron bands for the optimized structures under three different strain conditions are shown in Fig. \ref{band},
calculated within the local spin-density approximation with Coulomb interaction and spin-orbit coupling included (LSDA+SO+U). 
The Fermi surfaces for the doped SIO are shown in Fig. \ref{fermi-h}.
The results show systematic changes of the  band structure features in the gap region, some of which are shown in Fig. \ref{gap}. The magnitude of the fundamental gap $\Delta_g$ increases with tensile strain, and so does the direct gap $\Delta_X$ at the $X$ point, while the direct gap at the $\Gamma$ point does not change very much. 
This trend in the gap values may be expected, since tensile strain would reduce the band widths of the LHB and the UHB, while the Coulomb U is relatively unchanged. The opposite happens for the compressive strain. The $\Gamma$ point beats the trend because the valence top at
$\Gamma$ has a considerable contribution from the oxygen states. 
The approximately linear variation of the gap with strain is clearly visible in the partial density-of-states shown in Fig. \ref{pdos-strain}.
This trend of the gap variation with strain has already been noted in earlier works\cite{ZhangPRL13,LadoPRB015,Kimsr016}.

 {\it Strain induced $\Gamma$ - X crossover} --  An interesting feature of the band structure is the drastic change of the valence band maximum under strain, which would have a significant effect on the transport properties of the hole-doped structure. As can be seen from the band structure, Fig. \ref{band}, the conduction band minimum does not change under strain, always occurring at the M point, while it changes with strain. In the unstrained structure, DFT results show that the energy of the valence maximum at $X$ and $\Gamma$ are nearly the same. 
 From our calculations, the $\Gamma - X$ separation is $E_{\Gamma X} \equiv E_\Gamma - E_X = 0.01 $ eV, which compares well with the earlier DFT calculations in the literature, where $E_{\Gamma X}  \approx 0.01 - 0.07 $ eV\cite{KimPRL08,ZhangPRL13,LadoPRB015,PeitaoPRM18}.
Thus according to the DFT results, $\Gamma$ is slightly below $X$, which is opposite to the ARPES measurement \cite{YueNatcom16}, where the energy at $X$ point is found to be above $\Gamma$ by about 2 eV, i. e., $E_{\Gamma X}  \approx 0.2 $ eV \cite{YueNatcom16}.
For the tensile strain, the valence maximum at $\Gamma$ increases in energy, while that of $X$ goes down; For the compressive strain, the reverse happens, leading to a $\Gamma-X$ crossover with strain.

The $\Gamma - X$ crossover may be described within a tight-binding model involving the $|J_{\rm eff} = 1/2 \rangle $ orbitals on the square lattice of Ir atoms
and by including the dependence of the TB hopping integrals of the octahedral rotation angles which vary with strain.
This TB model is developed in \ref{sectionTB}.
The TB model shows the correct trend with the result:
  $\Delta E_{\Gamma X} (TB)  \approx 1.65  {\rm \ eV} \times  \  e_{xx}$ (Eq. \ref{DGX}) as compared to 
  $\Delta E_{\Gamma X} ({\rm DFT})  \approx 4.0 \ {\rm eV} \times \ e_{xx}$,  obtained from the DFT results shown in Fig. \ref{gap}.
  The TB analysis indicates that both the angle and distance changes with strain are important for the description of the $\Gamma-X$ crossover. 
  Based on the computed slope of $E_{\Gamma X}$ from DFT and the experimental $\Gamma-X$ separation of 0.2 eV, we would need 
an estimated epitaxial compressive strain of $e_{xx} \ge 3\%$ in order to switch the valence band maximum from $X$ to $\Gamma$.
 The $\Gamma-X$ crossover is clearly seen from the Fermi surface plots shown in Fig. \ref{fermi-h}, where we have shown the
 Fermi surfaces for 5\% dopant concentration. 

For the case of electron doping, we find that the conduction band minimum always occurs at the $M$ point in the 
Brillouin zone and the electron pocket is elliptical in shape. 

\subsection{Magnetic moment under strain}
\label{magnetic moment}

The calculated spin and orbital magnetic moments with and without strain are listed in Table \ref{table3}. 
We note that the total energy obtained with spin moments constrained along the planar direction ($\hat x$) has a lower energy than
the spin moments constrained along $\hat z$, consistent with the $\hat x$ orientation of the magnetic moments.
Table \ref{table3} shows the calculated magnetic moments using VASP with the LSDA+U+SO functional.

There has been much interest in the magnetic moments in SIO, in particular on the ratio  $\mu_l / \mu_s$ \cite{ZhangPRL13, LadoPRB015, Kimsr016, Haskel, Fujiyama, Franchini},  
since a ratio of two indicates the spin-orbital entanglement of the wave function and a deviation from this value is indicative of a mixture 
between $J_{\rm eff} = 1/2  $ and 3/2 sectors, as suggested from Table \ref{Table-Psi}.
For the unstrained material ($e_{xx} = 0$), the calculated total moment $  \mu_l + \mu_s  = 0.38 \mu_B$ is in good agreement with values obtained from  the magnetic susceptibility measurements ($0.5 \mu_B$) \cite{Cao} as well as  from previous calculations ($0.36 \mu_B)$\cite{KimPRL08}. 
As already mentioned, the $d^5$ configuration of Ir can be thought of as a single hole in the $t_{2g}$ manifold, leading to the cubic-field  values,  $\mu_l = 2/3 \  \mu_B$ and $\mu_s = 1/3 \  \mu_B$,  as seen from Table \ref{Table-Psi}.
 When a tetragonal field is present as in the case of SIO, the $| J_{\rm eff} , m \rangle$ states get mixed among themselves, and the 
 magnetic moments can be substantially altered from the cubic-field value. 
 
 From our calculations, we find $\mu_l / \mu_s \approx 2.6$ (Table \ref{table3}) for the unstrained structure, in general agreement with earlier calculations\cite{ZhangPRL13, LadoPRB015} as well as with a recent measurement 
 using non-resonant magnetic x-ray diffraction which obtained the value $\mu_l / \mu_s \approx 2.5$. 
 An earlier x-ray absorption measurement\cite{Haskel}  yielded the  ratio
$\mu_l / \mu_s \approx 1.1$; the reason for the discrepancy between the two measurements is unclear. 

The magnetic moment ratio as well as its variation with strain can be approximately described by invoking a tetragonal crystal field for 
the single ion in the presence of the spin-orbit coupling. For more accurate description,  
a renormalized spin-orbit coupling has been invoked\cite{ZhangPRL13}.
The results for the single ion in the tetragonal field are given in \ref{Appendix1}. For the spin moment aligned along $\hat x$, which is the case for SIO, we have 
$\mu_l / \mu_s = 2 + 4 \xi / 3$, where $\xi = \varepsilon / \lambda$ is the ratio of the tetragonal field to the spin-orbit coupling strength
$\lambda \approx 0.4$ eV. For the unstrained case, $\varepsilon \approx 0.14$ eV \cite{ZhangPRL13, KimNAT14},
so that the ratio $\mu_l / \mu_s = 2.47$,
in reasonable agreement with the DFT result stated in Table \ref{table3}.     
If we use the results of Ref. \cite{Franchini} which suggest that the magnitude of $\varepsilon$ increases (decreases) by about 0.1 eV for
tensile (compressive) epitaxial strain of 2 \%, then $\mu_l / \mu_s = 2.8$ for the tensile case and 2.1 for the compressive case, which more or less explains the DFT calculated trend shown in Table \ref{table3}.

\begin{table}[hb]   
\centering
\caption{The computed  spin ($\mu_s$) and  orbital ($\mu_l$) magnetic moments (in $\mu_B$) for the unstrained and strained structure.
}
\begin{tabular}{@{}l l l l}
\hline
\noindent$e_{xx}$&+2\% & 0 & -2\%\\
\hline

$\mu_s$  &0.084 & 0.108  &0.135\\ 
$\mu_l$ & 0.286  &  0.276  & 0.293\\  
$\mu_l/\mu_s$ & 3.4 & 2.6 & 2.2\\
\hline
\end{tabular}
 \label{table3} 
\end{table}


\section{Optical absorption}\label{opticsetc}

The change in the band structure with strain is reflected in the optical absorption spectrum.
The basic quantity to compute is the  imaginary part of the 
dimensionless
dielectric constant $\varepsilon_{2}(\omega)$, from which the real part $\varepsilon_1 (\omega)$ and the 
refractive index $n (\omega)$ are
computed using the Kramers-Kr\"onig relation. 
In the dipole-approximation, the optical absorption coefficient $\alpha (\omega)$  is  given by
\begin{equation}
\alpha (\omega) =    \omega  c^{-1}  n(\omega)^{-1} \times   \varepsilon_{2}(\omega) ,
\label{alpha}
\end{equation}
\begin{equation}
\varepsilon_{2}(\omega) =   \frac{4\pi^2e^2}{m^2\omega^2}\sum_{v, c}  \int_{BZ}  \frac {d^3 k} {( 2\pi)^3}      
| \hat e \cdot {\bf M}_{cv} ({\bf k})|^2       \delta(\hbar \omega-\epsilon_{c{{\bf k}}}+\epsilon_{v{{\bf k}}}),
\label{epsilon}
\end{equation}
where 
$ {{\bf M}}_{cv} ({{\bf k}})      = \langle \psi_{c{{\bf k}}}|  {\bf p} |\psi_{v {\bf k} }\rangle$
is the momentum matrix element between the conduction and the valence states (defined as unoccupied and occupied states, respectively), 
and $\hat e$ is the
light polarization vector. A closely related subsidiary function, useful to the discussion of the optical absorption, is the
joint density of states
\begin{equation}
{\rm JDOS} (\omega) =    \sum_{v, c}  \int_{BZ}  \frac {d^3 k} {( 2\pi)^3}      
     \delta(\hbar \omega-\epsilon_{c{{\bf k}}}+\epsilon_{v{\bf k}}).
\label{JDOS}
\end{equation}

It is important to note that even though the ${\rm JDOS} (\omega) $ scales up linearly with the size of the unit cell chosen,
the optical absorption coefficient $\alpha (\omega)$ and
$\varepsilon_{2}(\omega)$ are both independent of the size of the
unit cell used in the band calculation, as they must be. This is because a large number of transitions in the larger unit cell are simply forbidden.

This is easily seen by realizing that
 if a larger unit cell is used in the calculation, then a large number of vertical transitions in the folded Brillouin zone (larger unit cell)
is disallowed because the matrix element $M_{cv} ({{\bf k}})$ becomes zero from Bloch symmetry, 
even though the conduction and valence states both have nominally the same momentum ${\bf k}$.
Alternatively, one can start with the allowed optical transitions in the Brillouin zone corresponding to the smallest unit cell and then
fold them into the smaller Brillouin zone corresponding to the larger unit cell and explicitly see that not all vertical transitions are allowed in the smaller Brillouin zone. The joint density of states JDOS $(\omega)$, in contrast, scales with the size of the unit cell, which can be seen by computing its integral over energy by taking advantage of the $\delta$ function in the definition.
Note  that the dipole approximation used in the expressions omits the local field and excitonic effects. 
Results are presented below for plane-polarized light with  polarization direction in the plane  
or normal to the plane. 

\begin{figure}
\centering
\includegraphics[scale=0.5] {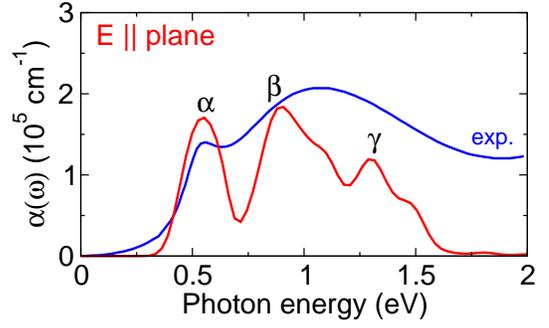} 
\caption{ Optical absorption coefficient $\alpha(\omega)$ for bulk SIO without strain and for light polarization
along the plane. The calculated spectra have been shifted to match with the measured optical gap. 
The origin of the three characteristic low-energy peaks $\alpha$, $\beta$, and $\gamma$
is indicated in  Fig. \ref{origin}. 
}
\label{absorption-bulk}
\end{figure}

\begin{figure*}
\centering
\includegraphics[scale=0.25]{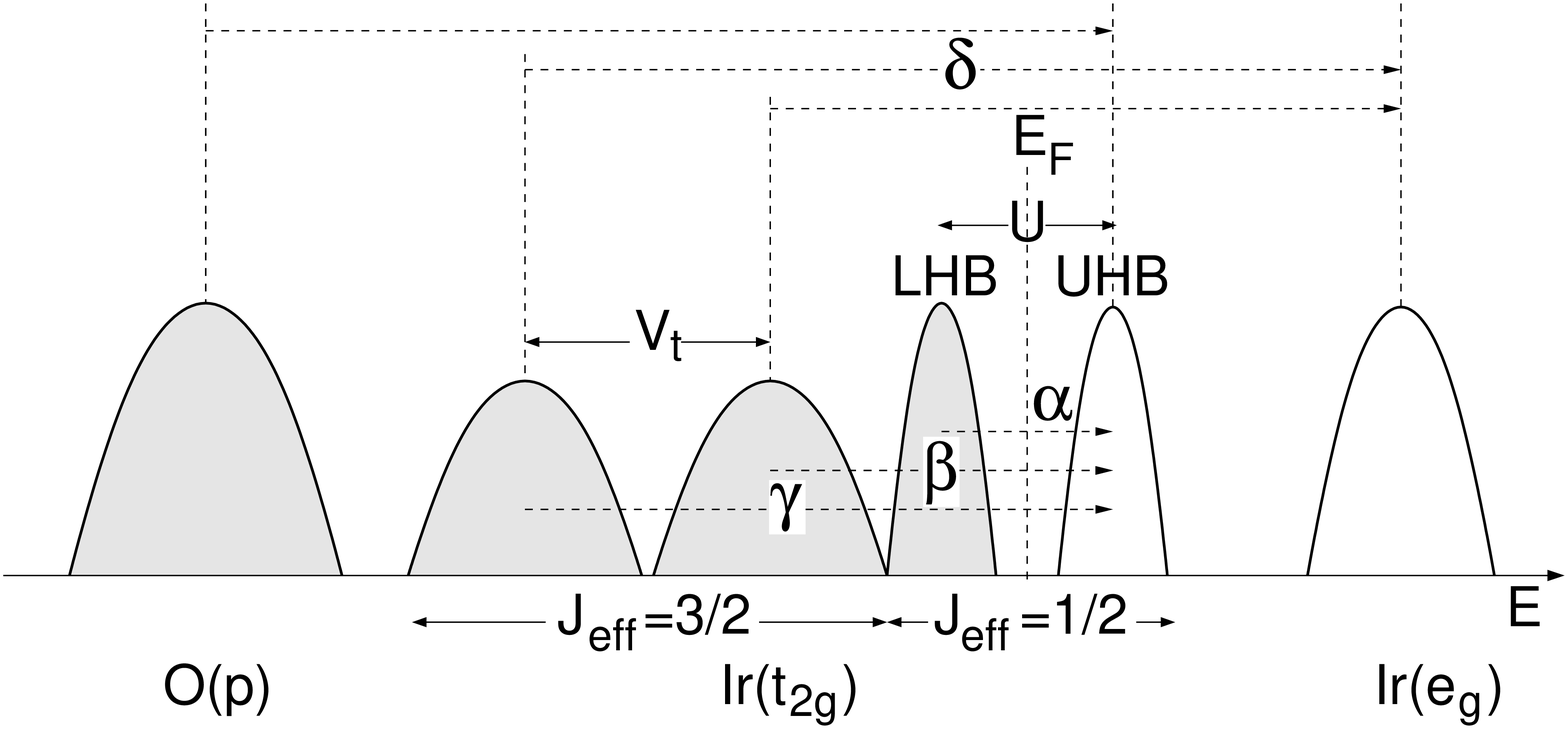} \\
\vspace*{1cm}
\includegraphics[scale=0.45]{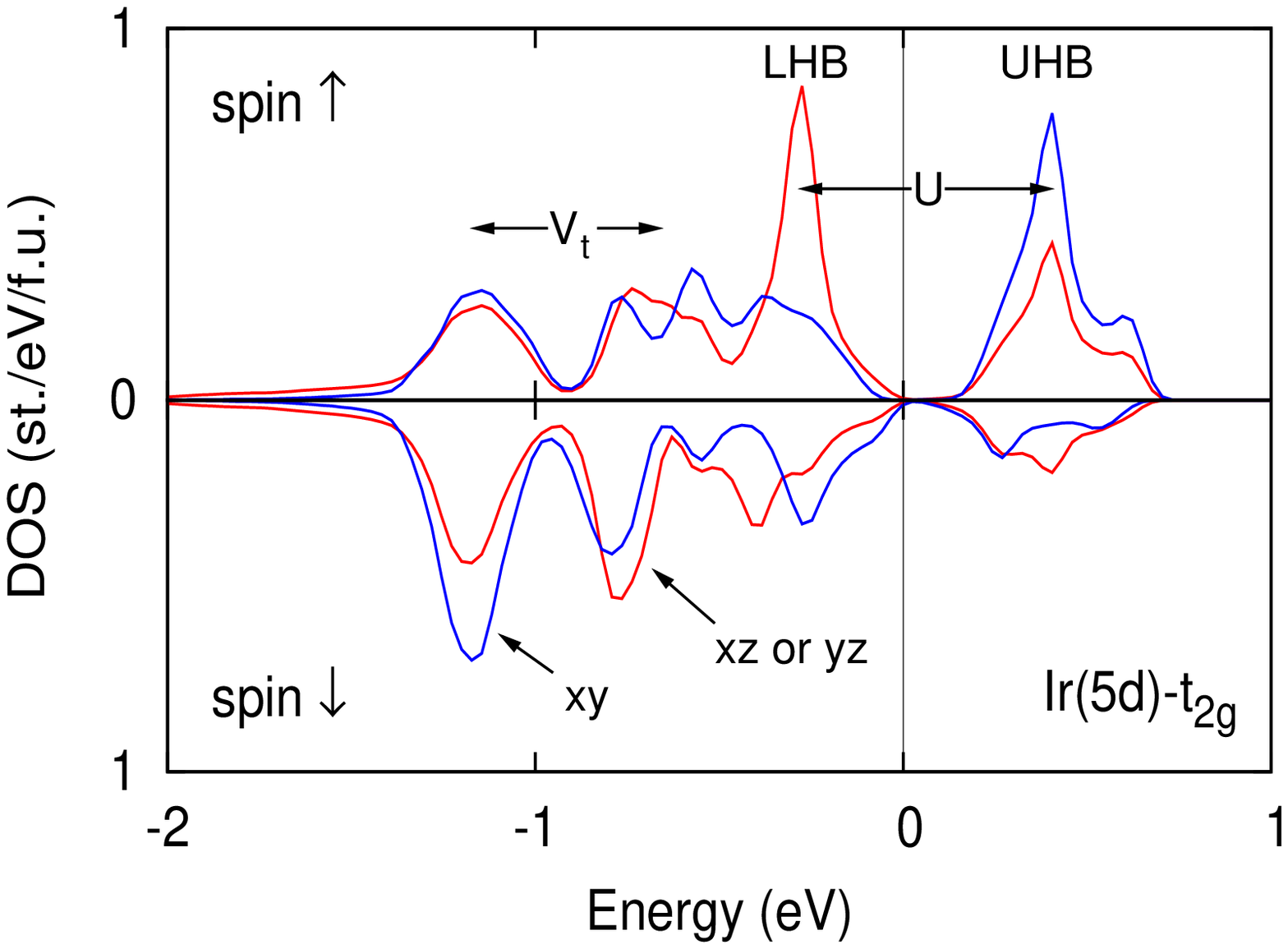} 
\caption{ Schematic orbital levels in SIO responsible for the optical transitions (top) and the Ir t$_{2g}$ levels from the DFT calculations (bottom).
The $J_{\rm eff} = 1/2 $ states are split by the Coulomb $U$ term, while the $J_{\rm eff} = 3/2 $ states are split due to the tetragonal crystal field $V_t$. The origin of the three peaks $\alpha, \beta$, and $\gamma$ and the $\delta$ transitions in the optical
absorption are indicated in the top figure. 
}
\label{origin} 
\end{figure*}

\begin{figure}
\centering
\includegraphics[scale=0.3]{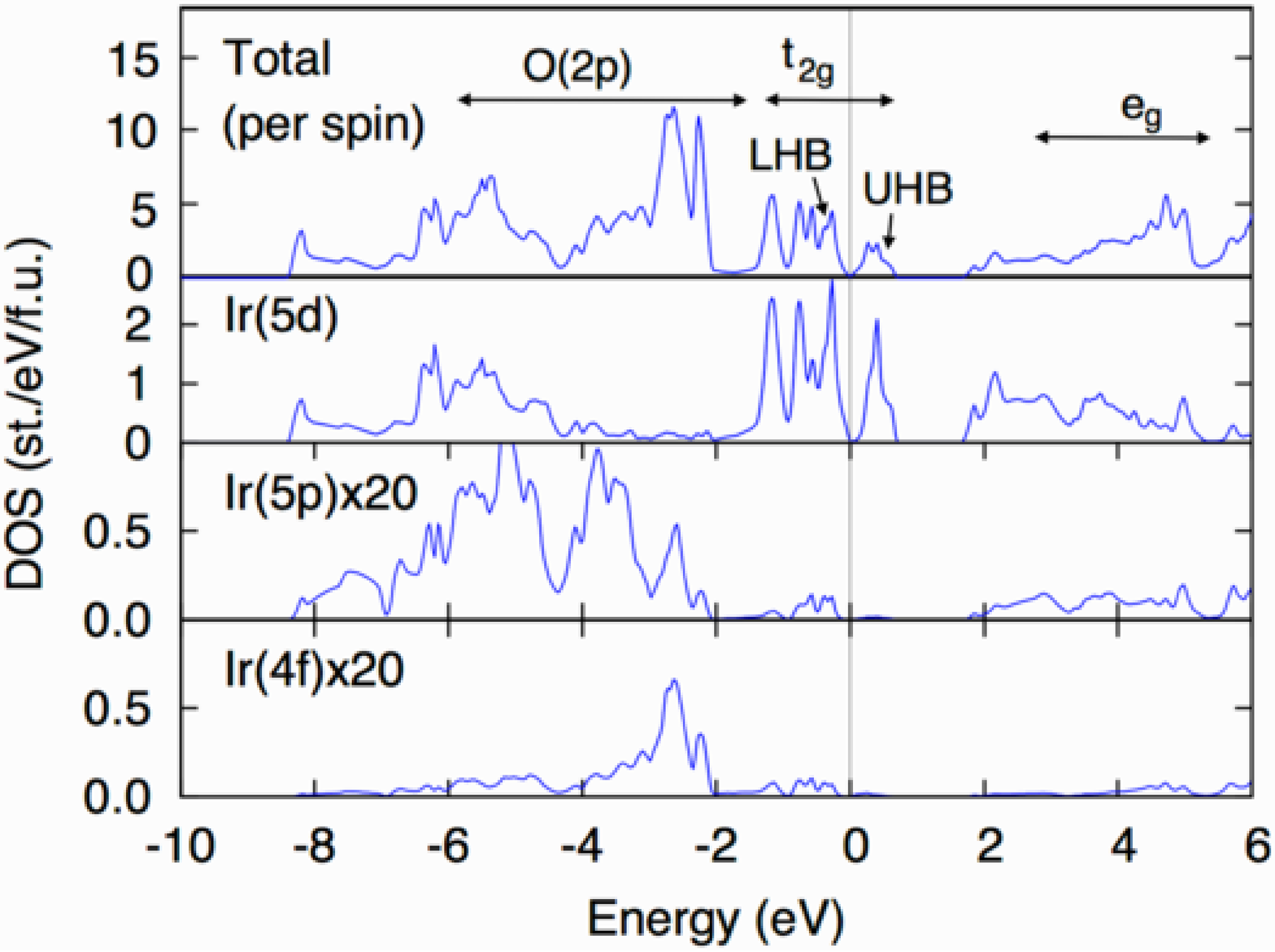}
\caption{ Admixture of the Ir (5p) and Ir (4f) orbitals into the Ir (5d) bands, which drives the $d-d$ optical transitions as discussed in the text.
The average admixtures of these orbitals in the energy range  - 2  to  5 eV, relevant for the Ir (d) bands are: $\eta^2 \sim 0.5\%$ for the Ir (5p) orbitals and 
$\eta_f^2 \sim 0.2\%$ for the Ir (4f) orbitals. 
}
\label{PDOS}
\end{figure}

\subsection{Unstrained bulk } 

The calculated absorption spectra for bulk SIO without strain are shown in Fig. (\ref{absorption-bulk}) for light polarization along the plane   ($E \parallel {\rm  plane}$), which is also compared to the measured data\cite{NicholsAPL13}.
The spectra show three distinct peaks which can be understood from the transitions between the 
$J_{\rm eff} = 1/2$ and $J_{\rm eff} = 3/2$ states as shown in Fig. (\ref{origin}). 
The $J_{\rm eff} = 3/2$ states are split into two as seen from Table \ref{table1} 
due to the tetragonal crystal field $V_t = 2\varepsilon / 3$.
In the experiment, only two distinct peaks are seen (these are labeled $\alpha$ and $\beta$), while the weaker $\gamma$ peak is missing, possibly
due to instrumental broadening. 

We did not find any significant asymmetry in $\alpha(\omega)$ for polarization directions within the plane. 
However, there is a large difference whether  $E \parallel {\rm  plane}$ or $E \parallel {\rm  z}$, as discussed later .

\subsection { Dipole selection rules and the d-d optical transitions} 

\begin{figure}
\vspace*{2cm}\includegraphics[scale=0.5] {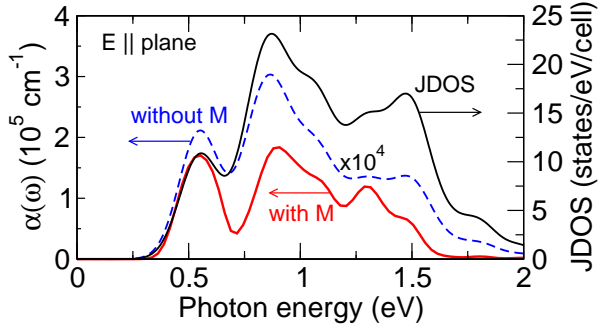} 
\caption{ Optical absorption coefficient calculated with and without  ($M_{cv} = 1$)  the matrix element  in
Eq.  (\ref{epsilon}) for unstrained SIO. 
In both cases, the same refractive index $n (\omega)$ in Eq. (\ref{alpha})   was used for a direct comparison. 
Since the square of the matrix elements enter into the calculation of $\alpha (\omega)$, the results show that
the average $M_{cv}^2 \sim 10^{-4}$ au.
The  optical joint density of states (JDOS) is computed for the unit cell of two formula units. 
\label{jdos}}
\end{figure}

The optical absorption in the low energy range occurs due to transitions within the Ir t$_{2g}$ manifold,
transitions that are however not 
dipole allowed due to the optical selection rules ($\delta L = \pm 1$ for plane-polarized light).
In this subsection, we discuss the optical transition matrix elements
and  conclude that they become dipole allowed due to the admixture of the Ir $p$ states into the Ir $d$ bands.
To get the dipole matrix elements, we need to estimate two things, as indicated from Eq. (\ref{matrix-element}) below: 
(i) The amount of the Ir p - d admixture $\eta$, which we estimate from perturbation theory as well as
from density-functional results and (ii) The optical matrix element for dipole-allowed transition between the Ir p and d orbitals
$\langle  p |  \hat e \cdot {\bf p} |d \rangle$, which we estimate from an effective hydrogenic model.

The d-d transitions are 
 non-zero in the crystal only because there is deviation from the spherical symmetry, the so called crystal field,
 which also produces splitting between the t$_{2g}$ and the $e_g$ states\cite{BhandariJPCS}. 
 The crystal field not only splits the Ir (5d) orbitals, but it also mixes  the Ir (5p) and (4f) orbitals into the 5d states, so that the d-d optical transitions have non-zero matrix element.
 Alternatively, such transitions may be equivalently described to be due to the transfer between Ir d atomic orbitals to the adjacent atoms because of the wave function overlap, e. g., within the linear combination of atomic orbitals (LCAO) model.

We now estimate the dipole matrix elements, which are non-zero due to the $p-d$ and $f-d$ mixing for the Ir atom. 
We first consider the $p-d$ mixing, which as we shall see makes the larger contribution.
Symbolically, this is
given by
\begin{eqnarray}
M_{dd}     &=&   \langle \psi_{c {\bf k}}|  \hat e \cdot {\bf p} |\psi_{v {\bf k} }\rangle    \nonumber \\
&=&
\langle d + \eta p |  \hat e \cdot {\bf p} |d + \eta p\rangle
\approx      2 \eta  \times \langle  p |  \hat e \cdot {\bf p} |d \rangle.
\label{matrix-element}
\end{eqnarray}
The admixture $\eta$ of the Ir $p$ states into the Ir $d$ bands, although small due to the large $p-d$ energy difference, is nevertheless significant enough and is largely responsible for the optical transition in  SIO.

The magnitude of $\eta$ can be estimated from the partial density of states (PDOS) in the band calculations in the energy range of the Ir $d$ bands. With the definition for the PDOS,
$\rho_d (\varepsilon) = \sum_i | \langle \psi_i | d \rangle |^2 \delta (\varepsilon - \varepsilon_i)$ and  
$\rho_p (\varepsilon) = \sum_i | \langle \psi_i | p \rangle |^2 \delta (\varepsilon - \varepsilon_i)$, 
and with the wave functions 
$\psi \approx d + \eta p$, one can estimate the average value of the admixture from the two PDOS in the energy range of interest. 
The PDOS, computed using the LAPW method,\cite{LAPW} are shown in Fig. \ref{PDOS} and we have obtained the value of $\eta$ from the ratio
of the integral of the two PDOS  in the energy window of the Ir t$_{2g}$ bands, viz., from -2 to 1 eV. 
The result is: $\eta^2 \approx \rho_p / \rho_d \approx 0.19 \times 10^{-2}$ or $\eta \approx 4\%$. We find a similar admixture of the $4f$ states, $\rho_f/\rho_d \approx 0.2 \times 10^{-2}$, indicating a nearly equal $4f$ admixture $\eta_{4f} \approx 4\%$ into the Ir (5d) bands.

Alternatively, the strength of the admixture $\eta$ can be estimated from perturbation theory. From the second order  perturbation theory, we have
\begin{equation}
\eta = 3 \times  \langle p | V_{cf} | d \rangle  / ( \varepsilon_d - \varepsilon_p ), 
\label{eta}
\end{equation}
where the factor of three is from the degeneracy of the $p$ states. 
If we further approximate  $\langle p | V_{cf} | d \rangle  \sim \langle d | V_{cf} | d \rangle \approx \Delta_{cf}$, admittedly a crude approximation, and take the  t$_{2g}$ -  e$_g$ splitting $\Delta_{cf} \sim 2$ eV, and use the result  $\varepsilon_d - \varepsilon_p \sim 43$ eV for Ir from the standard Atomic tables\cite{Herman}, we get the perturbation theory result of  $\eta \approx 14 \%$, which is a factor of three too high as compared to the DFT result. However, given the crudeness of the approximation we made for the matrix elements in the perturbation theory, the order of magnitude agreement is reasonable. In the following, we shall use the DFT value for the $p-d$ admixture  $\eta \approx 4 \%$.

The second part is the estimation of the 
dipole matrix element $M_{pd} \equiv \langle  p |  \hat e \cdot {\bf p} |d \rangle$ appearing in Eq. (\ref{matrix-element}), which we do  
from an effective hydrogenic  model\cite{ZZ, Pauling}.
In this model, the atomic wave functions are described as hydrogenic wave functions, but with an orbital dependent effective nuclear charge $Z$.  Thus
\begin{eqnarray}
\Psi_{nlm} &=& R_{nl} (r) Y_{lm} (\Omega), \nonumber \\
R_{nl} (\rho) &=&  N_{nl} (Z) \rho^l e^{-\rho/2} L_{n+l}^{2l +1} (\rho) ,
\end{eqnarray}
where $\rho = (2Z/na_0) r$. Note that $Z$ here is an effective atomic number, which takes into account the 
screening of the core electrons and depends on the principal quantum number $n$ also. We take the value\cite{ZZ} $Z = 18.7$ appropriate for Ir
$n=5$ orbitals (5p and 5d).
 The integration can be performed analytically to yield the result $M_{pd}$ for the plane polarized light:
\begin{equation}
M_{pd} = \langle  \Psi_{510} |  p_z |\Psi_{520} \rangle = - 0.25 i
\end{equation}
in atomic units (viz., $\hbar = 1$, Bohr radius $a_0 =1$, $m_e = 1/2$, Energy unit = 1 Ryd). Note that for atoms, the direction of polarization doesn't matter due to spherical symmetry. Plugging in the estimated magnitudes of $M_{pd}$ and $\eta $  in Eq. (\ref{matrix-element}),
we find the $d-d$ matrix element due to the admixture of Ir (5p) into the Ir (5d) states
\begin{equation}
|M_{dd}^p|  = 2\eta \times |M_{pd}| = 2 \times 4\% \times 0.25 = 2 \times10^{-2}.
\label{M} 
\end{equation}

A similar calculation for the matrix element due to the admixture with the Ir (4f) orbitals yields a number, which is an order of magnitude
smaller than $|M_{dd}^p|$, viz.,
\begin{equation}
|M_{dd}^f| = 2 \eta_f \times |M_{fd}| = 2 \times 4\% \times 0.064 = 5 \times10^{-3},
\end{equation} 
where we estimated the 4f admixture $\eta_f$ from Fig. (\ref{PDOS}) and used the effective atomic number $Z = 38.3$ for the Ir 4f orbitals\cite{ZZ}.  
Thus the Ir (4f) orbitals contribute a much smaller amount 
as compared to the Ir (5p) orbitals. 

The estimate  given by  Eq. (\ref{M}), $M_{dd} \sim 10^{-2}$,  is very comparable to the average matrix element $M_{cv} \sim 10^{-2}$  obtained from the band calculations (Fig. \ref{jdos}), indicating that the admixture of the Ir (5p) orbitals is indeed responsible for the optical absorption. 

\begin{figure}
\includegraphics[scale=0.6] {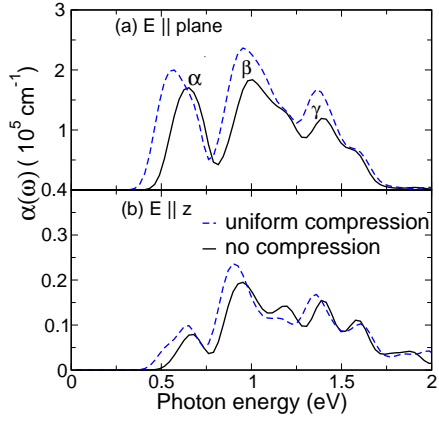} 
\caption{ Calculated optical absorption  under {\it uniform} compression:
(i) no compression  (solid lines) and (ii) uniform compression ($e_{xx} = e_{zz} =  - 2 \%$)
 (dashed lines). 
All atom positions were scaled without any structural relaxation.
}
\label{bulk-compression} 
\end{figure}

\begin{figure}
\includegraphics[scale=0.6] {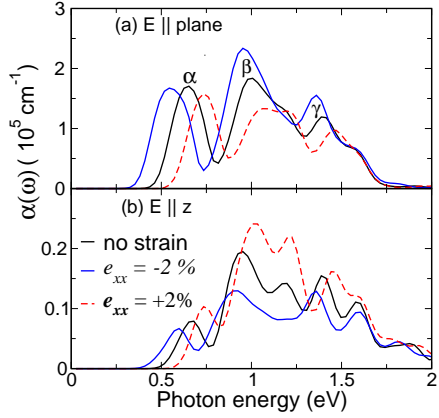}
\caption{ Calculated optical absorption spectra for different epitaxial strains. }
\label{optical-strain}
\end{figure}

\subsection{Optical absorption under strain}\label{optics}

{\it Uniform strain} --
The calculated optical absorption for uniform compression (uniform pressure) is compared to the same for bulk SIO without any compression in Fig. (\ref{bulk-compression}). Apart from a shift in the peak positions to lower energies, caused by a reduction in band gap due to the
increase of the band width due to compression, we note that the overall absorption coefficient is increased for both light polarizations.
This can be explained from our above argument of Ir 5p - 5d admixture, because a larger crystal field upon compression mixes the Ir 5p more
into the Ir 5d bands, making the optical matrix element larger. 
Indeed, following the same logic, as the lattice constant is increased, there is less and less admixture of the Ir (5p) orbitals, and in the limit of infinite lattice constant, the admixture vanishes (perfect spherical symmetry), so that optical absorption would be zero as would be expected for the $d \rightarrow d$  transition due to the dipole selection rules.

{\it Epitaxial strain} --
The change of the optical absorption with epitaxial strain is shown in Fig. (\ref{optical-strain}). 
With compressive epitaxial strain ($e_{xx} < 0$ and $e_{zz} > 0$), the distances in the plane are reduced, leading  to a larger optical absorption for E $\parallel$ plane  as argued for the uniform compression case.
The same effect leads to an overall reduction of the optical absorption for E $\parallel \hat z$. 
For the absorption with E $\parallel$ plane,
all three peaks $\alpha$, $\beta$, and $\gamma$ are still there, 
but the peaks are red (blue) shifted with compressive (tensile) epitaxial strain, consistent
with the band structure changes under strain shown in Fig. (\ref{band}).

\begin{figure}
\includegraphics[scale=0.50] {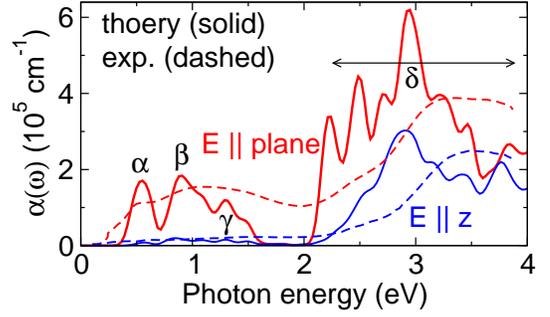}
\caption{ Polarization dependence of the absorption spectrum, compared to the experiments. Theory results are shown as full lines, while
dashed lines indicate the measurements of Nichols et al.
\cite{Nichols2-APL13}.
}
\label{optical-polarization}   
\end{figure}   

\begin{figure}
\includegraphics[scale=0.25] {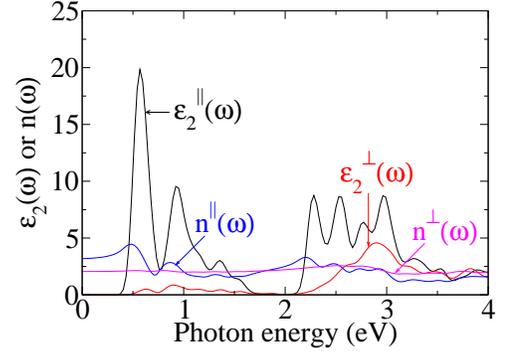} 
\caption{ Polarization dependence of the imaginary part of the dielectric constant $\epsilon_2 (\omega)$ 
and the refractive index $n (\omega)$. The superscripts $\parallel$ and $\perp$ indicate light polarization parallel 
(E $\parallel$ plane) and perpendicular  (E $\parallel \hat z$) to the plane, respectively.
}
\label{Fig-n-epsilon}  
\end{figure}  

{\it Optical anisotropy} -- 
The polarization dependence of the optical absorption has been reported in the literature\cite{Nichols2-APL13}, 
where a significant anisotropy is found in the optical spectra. The absorption is significantly reduced for the  polarization
E $ \parallel \hat z$ as compared to E $\parallel$ plane in the low energy region 
corresponding to the transition within the Ir t$_{2g}$ manifold, while for the higher energy region, the difference in the relative strength is not as drastic.
This is reproduced quite well from our calculations as seen from Fig. \ref{optical-polarization}, where we also compare with the existing experiments. 
The calculated spectra is for the unstrained structure, while the experimental spectra is for a system with non-uniform strain,
the only case
for which experimental results are available for both polarizations.
The anisotropy is directly attributable to the differences in the matrix element $| \hat e \cdot {\bf M}_{cv} ({\bf k})|$ (Eq. \ref{epsilon}),
since the variation of the refractive index $n (\omega)$ with energy or polarization is relatively weak, as indicated from Fig. \ref{Fig-n-epsilon}.

\section{Summary}      \label{con}

In summary, we studied the electronic properties and the optical absorption spectra of Sr$_2$IrO$_4$ under epitaxial strain condition using density-functional methods. Systematic structural changes with strain were found including the staggered rotation angle $\theta$, which was found to have important effect on the electronic structure. An interesting result is the $\Gamma - X$ crossover of the $J_{\rm eff} = 1/2 $ valence band maximum with strain, allowing for engineering of the hole pocket in the hole-doped material, with potentially drastic changes in the transport properties. 
A minimal tight-binding Hamiltonian was developed for the $J_{\rm eff} = 1/2 $ sector, which is capable of qualitatively describing the
important features of the band structure under strain, including the $\Gamma- X$ crossover.

We calculated the optical spectra under strain, interpreting the results in terms of small admixtures of the Ir (5p) and (4f) states with the Ir (5d) bands, without which the optical transitions would be dipole forbidden. The calculated spectra were compared to the experiments, where available, and the observed anisotropy in the optical absorption, which is very strongly anisotropic for the low energy region but 
less strongly anisotropic for the high energy region (see Fig. \ref{optical-polarization}), was correctly  
explained from the calculated results.
Our work opens up the possibility of exploring the strain manipulation of the transport properties of epitaxially grown spin-orbit coupled Mott systems.


\acknowledgements{We thank Jamshid Moradi Kurdestany for valuable discussions 
and the U.S. Department of Energy, 
Office of Basic Energy Sciences, Division of Materials Sciences and Engineering (Grant  No. DE-FG02-00ER45818) for financial support. Computational resources were provided by the National Energy Research Scientific Computing Center, a User Facility also supported by the U.S. Department of Energy. 

\appendix

\section{Tight-binding model and the $\Gamma - X$ crossover } \label{sectionTB}
In this Appendix, we construct a minimal tight-binding model to describe the band structure within the $J_{\rm eff} = 1/2$ sector on the square lattice
of Ir atoms, appropriate for SIO, in order to understand the $\Gamma - X$ crossover of the valence band with epitaxial strain.


We are primarily interested in the $J_{\rm eff} = 1/2$ sector, viz., the lower and the upper Hubbard bands (LHB / UHB)  that form the valence and the 
conduction bands in the gap region. 
A tight-binding description can follow two different paths: One is to keep the $t_{2g}$ orbitals (in total six orbitals in the basis per Ir atom including spin) in the Hamiltonian and then fit the the two Hubbard bands in the DFT band structure and the other is to keep only the two $J_{\rm eff} = 1/2$  orbitals per Ir atom (the minimal model). 
For the fitting with the $t_{2g}$ orbitals, the NN TB parameters that best fit the two Hubbard bands are (in eVs): $V_\pi = -0.20, V_\delta = 0.04, U = 0.78, \lambda = 0.4$. 
In our discussions below, we employ the second approach, where we use the minimal  TB model with just the two orbitals per Ir atom to describe the $J_{\rm eff}  = 1/2$ sector.

We consider the Hubbard model on a square lattice with anti-ferromagnetic order as appropriate for SIO, 
keeping the two spin-orbital entangled $J_{\rm eff}  = 1/2$ orbitals on each site, 
which we call $ e_1$ and $e_2$, defined with respect to the 
{\it local} octahedral axes, with the staggered rotations as indicated in Fig. (\ref{rotation}), viz.,
\begin{eqnarray}
\label{e1-e2}
|e_1\rangle &\equiv& |\frac{1}{2}, -\frac{1}{2} \rangle =
(|xy\uparrow\rangle + |yz\downarrow\rangle + i|xz\downarrow\rangle)/\sqrt 3 \nonumber\\
|e_2 \rangle&\equiv& |\frac{1}{2}, \frac{1}{2} \rangle =
(|yz\uparrow\rangle - i|xz\uparrow\rangle - |xy\downarrow\rangle)/\sqrt 3.
\end{eqnarray}
The TB Hamiltonian is  given by
\begin{equation}
{\cal H} = \sum_{ \langle ij\rangle   \alpha} t_{ij} c^\dagger_{i\alpha}c_{j\alpha}+h.c. + \frac{U}{2} \sum_{i\alpha} n_{i\alpha} n_{i\bar \alpha},
\end{equation}
where $c^\dagger_{i\alpha}$ creates an electron at the site $i$ (which may be in sublattice A or B) and in the orbital   $e_\alpha$, $t_{ij}$ is the hopping integral, which is non-zero only for hopping between the same type of orbitals $e_1$ or $e_2$, $U$ is the on-site Coulomb interaction, and the summation $\langle ij\rangle$
indicates sum over distinct pairs of bonds.  Note that in general the hopping integral $t_{ij}$ are complex numbers as discussed below.
We will retain hoppings up to the fourth NN as indicated  in Fig. (\ref{fig-hopping}). 
Furthermore, we find that at least three (and sometimes four) NN hoppings need to be kept  for an accurate description of the band structure in this minimal
model for the  $J_{\rm eff}  = 1/2$ sector.

To obtain the hopping integrals between these orbitals, it is convenient to first obtain the integrals in the {\it unrotated} $d$ basis from standard Tables\cite{Harrison}
and then rotate the basis.
Under rotation the angular momentum functions transform among one another, and in the present case, we have a 
 site-dependent $5 \times 5$ rotation matrix appropriate for $L=2$.
We denote the unrotated basis as $|\alpha \rangle$ ($ xy, yz, zx, x^2 - y^2,$ and $3z^2 -1 $, in that order), rotated basis by 
$|\alpha^\prime \rangle$ on sublattice A and 
$|\alpha^{\prime \prime} \rangle$ on sublattice B, 
and the corresponding rotation matrices that
transform one basis into another  by 
$R^\prime$ and $R^{\prime \prime}$,  i. e., $|\alpha^\prime \rangle = R^\prime |\alpha\rangle$ and
$|\alpha^{\prime \prime}  \rangle = R^{\prime \prime} |\alpha\rangle$.
The hopping integrals in the rotated basis are then given by  
$\widetilde H_{\alpha^\prime \beta^{\prime \prime}} \equiv    \langle \alpha^{\prime } | H | \beta^{\prime \prime} \rangle = 
\langle \alpha | R^{\prime T} HR^{\prime \prime} |\beta \rangle$, or
\begin{equation}
\widetilde {H} =  R^{\prime T}  H R^{\prime \prime}. 
\label{H-tilde}
\end{equation}
The rotation matrix  $R (\theta)$ for $L =2$ with rotation $\theta$ about the $\hat z$  axis is well known\cite{Avinash-osmate}

 \begin{eqnarray}                
 R (\theta) =       \begin{pmatrix}
\cos2\theta & 0 & 0 & \sin2\theta & 0\\          0 & \cos\theta & \sin\theta & 0 & 0\\
 0 & -\sin\theta & \cos\theta& 0 & 0\\         -\sin2\theta & 0 & 0 & \cos2\theta & 0 \\         0 & 0 & 0 & 0 &1 
\end{pmatrix}.   
\end{eqnarray} 
Due to the staggered rotations, the hopping integrals between orbitals on the same sublattice and opposite sublattices are given using Eq. (\ref{H-tilde}) as follows:
$ \widetilde {H}_{AA} =  R (\theta)^{ T}  H_{AA} R (\theta)$, $ \widetilde {H}_{AB} =  R (\theta)^{ T}  H_{AB} R (-\theta)$, $ \widetilde {H}_{BB} =  R (-\theta)^{ T}  H_{BB} R (-\theta)$, and $ \widetilde {H}_{BA} =  R (-\theta)^{ T}  H_{BA} R (\theta)$. 
Denoting the two rotated $J_{\rm eff} = 1/2$ states corresponding to Eq. (\ref{e1-e2}) by $|\widetilde e_{1A} \rangle$, $|\widetilde e_{1B} \rangle$, $|\widetilde e_{2A} \rangle$, and $|\widetilde e_{2B} \rangle$, one can then obtain the hopping integrals between these set of orbitals from the standard $d-d$ hopping integrals from Harrison's Tables\cite{Harrison} and using the above rotation matrices. 
With the standard notations for the direction cosines $nlm$ for the distance vector joining the first atom to the second, these matrix elements are readily obtained. The hopping amplitude between the opposite sublattice is 
\begin{eqnarray}
&&\langle \widetilde e_{iA}| H | \widetilde e_{j B} \rangle = h_i \delta_{ij},  \nonumber \\
&& h_1 = t + i t^\prime, \ \ h_2 = h_1^*,  \nonumber 
\end{eqnarray}
\begin{eqnarray}
 3t &&=[  3l^2m^2V_{\sigma}+(1-4l^2m^2)V_{\pi}  ]  \cos^22\theta  \nonumber \\ 
& &
 - [ \frac{3}{4}(l^2-m^2)^2V_{\sigma}+ (1-(l^2-m^2)^2 )  V_\pi ]   \sin^2   2\theta  \nonumber \\
 &&  + V_{\pi} \cos 2\theta,  
\label{hopping-opposite}
\end{eqnarray}
\begin{eqnarray}
t^\prime = -   \frac{ V_\pi} {3}  \sin2\theta.\nonumber
\end{eqnarray}
Notice that the hopping matrix is diagonal with complex elements for non-zero $\theta$, which however can be made real by making a 
gauge transformation. For hopping between the same sublattice, the hopping matrix is diagonal, but with real elements this time:
\begin{eqnarray}
&&\langle \widetilde e_{iA}| H | \widetilde e_{jA} \rangle = t \delta_{ij},  \nonumber \\
&& 3t =  V_\pi + [ V_\pi -(l^2-m^2)^2   
(V_\pi- 3    V_\sigma / 4) ]   \sin^2   2\theta    \nonumber \\
&& + [ V_\pi-l^2m^2(4V_\pi-3V_\sigma) ]  \cos^2 2\theta    \nonumber \\
&& lm (l^2-m^2) (2V_\pi- 3 V_\sigma / 2)\sin4\theta.
\label{hopping-same}
\end{eqnarray}
We make a further simplification by taking $V_\sigma = -(3/2 )V_\pi$ following Harrison's scaling\cite{Harrison}  in order to construct a minimal model.

The same expression Eq. (\ref{hopping-same}) is valid for hopping between B sublattice atoms $\langle \widetilde e_{iB}| H | \widetilde e_{jB} \rangle$, except that 
the sign of $\theta$ is changed due to the staggered rotations. 
Sometimes, in the literature, a simplistic angle dependence for the hopping $h (\theta) = h_0 \cos \theta$ is used. 
However, as the two equations above demonstrate, the angle dependence is more complicated and cannot be written down as $\cos \theta$ even to
the lowest order in the angle.
In our TB model, we have retained up to four NN hoppings, as indicated in Fig. \ref{fig-hopping}. 
 
\begin{figure}
\centering
\includegraphics[scale=0.40]{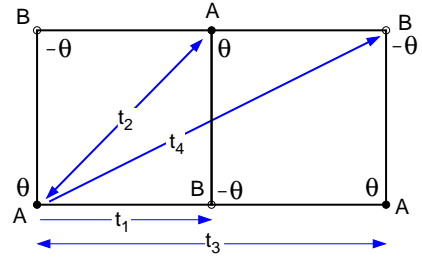}
\caption{ Hopping integrals between the $| \widetilde e_1 \rangle$ orbitals, which can be obtained from 
Eqs. (\ref{hopping-opposite}) and (\ref{hopping-same}). The same-sublattice hoppings $t_2$ and $t_3$ are 
$\theta$ independent and real,
while the opposite-sublattice hoppings, $t_1$ and $t_4$, depend on $\theta$ and are complex if $\theta \ne 0$. Hopping integrals for $| \widetilde e_2 \rangle$ orbitals are complex conjugate of those for the
$| \widetilde e_1 \rangle$ orbitals.
}
\label{fig-hopping} 
\end{figure}

{\it Gauge transformation} -- As seen from Eq. (\ref{hopping-opposite}), the hopping integrals are in general complex,
and of course there is no problem working with the complex hoppings. 
However, it is convenient to make the NN integral real by a gauge transformation\cite{Avinash-2017}. 
In the transformation, one simply adds a multiplicative phase factor to the orbital definitions, viz.,
\begin{eqnarray} \label{gauge}
&& |\widetilde e_{1j}   \rangle \rightarrow          |\widetilde e_{1j} \rangle       e^{i \varepsilon_j \phi /2 } \nonumber \\
&& |\widetilde e_{2j}   \rangle \rightarrow   |\widetilde e_{2j}    \rangle         e^{-i \varepsilon_j \phi /2},
\end{eqnarray}
where  $\varepsilon_j = \pm 1$ for $ j = A/ B$ sublattices. With the choice $\phi = \tan^{-1} (t^\prime /t)$, where $t, t^\prime$ are 
given in  Eq. (\ref{hopping-opposite})
for the 1NN hopping,
the new hopping integral becomes real, viz.,  $ t +i t^\prime \rightarrow \tilde t$, where $\tilde t = (t^2 + t^{\prime 2})^{1/2} $. 
This transformation leaves  all same-sublattice hoppings real, while the opposite-sublattice hoppings including $t_4$ and beyond continue to remain complex. One can choose to work with such complex hoppings, or else, simply ignore the imaginary parts, since hoppings between far-away neighbors are small anyway. We denote the gauge-transformed basis (\ref{gauge}) as
$|e_1\rangle$ and $|e_2\rangle$ in Eq. (\ref{e1-e2})  and the 
 hopping amplitudes (real)   as $t_i$ as indicated in Fig. (\ref{fig-hopping}).

The antiferromagnetic lattice structure together with the fact that $|\widetilde e_1\rangle $ and $|  \widetilde e_2\rangle $ subspaces do not mix leads
to the $2 \times 2$ TB Hamiltonian in the momentum space
\begin{equation}
H (\bf k) = \left(\begin{array}{cc} 
-\Delta + h_{11}  & h_{12} \\ 
h_{12}^* & \Delta + h_{11} 
\end{array}
\right),
\label{H22}
\end{equation}
where $h_{11} = 4t_2  \cos k_x \cos k_y + 2 t_3 (\cos 2 k_x + \cos 2 k_y) $, 
$h_{12} = 2 t_1 (  \cos k_x + \cos k_y)  + 4 t_4 (\cos 2 k_x  \cos  k_y + \cos  2 k_y  \cos   k_x) $, $a_0 = 1$ here, and $\Delta = U/2$ is the staggered field. Diagonalization readily yields the energies of the upper and lower $J_{\rm eff} = 1/2$ Hubbard bands to be
\begin{equation} 
\label{TB-Energy}
\varepsilon_\pm ({\bf k}) = h_{11}  ({\bf k}) \pm \sqrt { \Delta^2 + h_{12}^2  ({\bf k}) }.
\end{equation}

The TB parameters obtained from fitting Eq. (\ref{TB-Energy}) to the density-functional bands for the optimized structure with no strain
are, in units of eV: 
$U = 0.65$, $t_1 = -0.095$, $t_2 =0.015$, $t_3 = 0.035$, and $t_4 = 0.01$. 
The TB fit to the DFT bands are shown in Fig. (\ref{band}) as 
dotted lines. An important feature of the band structure is the occurrence of the conduction minimum at the $M$ point of the Brillouin zone, which the electrons would occupy in the electron doped system. It can be easily shown that for this to happen, the
TB parameters must satisfy the condition\cite{Sayantika}
\begin{equation} \label{condition}
t_3 > t_2/2 > -t_1^2/U,   
\end{equation}
which is clearly satisfied by our parameters given above.
Furthermore, the eccentricity of the elliptical energy contours around the  $M$ point in the conduction band,
as seen in Fig. \ref{fermi-h}, is given by\cite{Bhandari18}
  $ e  \equiv (1 - r^2)^{1/2}$, where the axis ratio 
$r = [ (2 t_3 - t_2) /  (t_2 + 2 t_3 + 4 (t_1-2t_4)^2 / U)   ]^ {1/2} \approx 0.6$ for SIO.

We determine the TB parameters under strain condition with the following ansatz. 
Using the calculated angle $\theta \approx 13^\circ $ and the hopping expressions Eqs. (\ref{hopping-opposite}) and (\ref{hopping-same}), there is a one to one correspondence between $t_i$ and $V_\pi^i$ for the $i$-th neighbor.
 We obtain the values of $V_\pi^i$ for different neighbors under no strain condition. In order to compute the hopping integrals under strain conditions, 
 we back substitute  $V_\pi^i$ and the DFT optimized angles under strain into Eqs. (\ref{hopping-opposite}) and (\ref{hopping-same}).
In addition to the change of the angles, distances between atoms also change.  
Taking the variation of $V_\pi$ to follow Harrison's
$R^{-5}$ scaling with distance $R$  and including the effect of $\theta$ from Eqs. (\ref{hopping-opposite})  and (\ref{hopping-same}),  we can compute the TB hopping parameters $t_i$ under the epitaxial strain conditions. 
We assume that strain enters the hopping parameters only via change of the hopping distances and the rotation angle $\theta$.

With this simple TB model, we find that the important trends of the 
band structure under strain are described qualitatively correctly,
though not quantitatively with a factor of two to three discrepancy as compared to the DFT results. 

In particular, the model can qualitatively describe the $\Gamma - X$ 
crossover of the valence band top under strain (hole pocket in the doped structure), predicted from the DFT calculations. 
From the TB energy expression Eq. (\ref{TB-Energy}), we readily find the $\Gamma - X$  energy difference for the valence 
band
\begin{equation} \label{Gamma-X}
E_{\Gamma X} = \varepsilon_{-} (\Gamma) - \varepsilon_{-} (X) = 8 t_2 + \Delta - [ \Delta^2 + 16 (t_1+2 t_4)^2 ]^{1/2}.
\end{equation}
For our TB parameters, this is almost zero consistent with the DFT results. 
The strain dependence of  $E_{\Gamma X}$ is obtained from the TB expression (\ref{Gamma-X}), viz.,
\begin{equation} \label{DGX}
\Delta E_{\Gamma X} = \frac{\partial   E_{\Gamma X} }  {\partial R} \Delta R +   
\frac{\partial   E_{\Gamma X} }  {\partial \theta} \Delta \theta  \approx 1.65  {\rm \ eV} \times  \  e_{xx},
\end{equation}
where we have used the fact that the distance change between the atoms in the plane is simply
$\Delta R / R = e_{xx}$,
and  also the DFT result for the change in angle with strain, viz., 
  $\Delta \theta  \approx (-1.5 \ {\rm rad} ) \ e_{xx}$, obtained from Table \ref{table1} assuming a linear strain dependence,
  since the strain is small. It turns out that the two terms in Eq. (\ref{DGX}) contribute nearly half each to the final result, so that both the angle and distance changes are important for the description of the $\Gamma-X$ crossover.

\section{Single ion in a tetragonal field with spin-orbit coupling} 
\label {Appendix1}

It is instructive to examine the eigenstates of the single site Hamiltonian for the $d$ orbitals in the presence of the SOC and a tetragonal crystal field, which this Appendix deals with.

We assume a large cubic crystal field splitting $\Delta_{cf} \rightarrow \infty$, so that the $e_g$ and the $t_{2g}$ sectors don't mix. Within the  $t_{2g}$ sector, the Hamiltonian is given by
\begin{widetext}
\begin{equation}
H_{SOC}=
\bordermatrix{~& |xz\uparrow\rangle & |yz\downarrow\rangle & |xy\downarrow\rangle & |xz\downarrow\rangle & |yz\uparrow\rangle & |xy\uparrow\rangle\cr
\langle xz\uparrow| & 0 & -i\lambda/2 & i\lambda/2 & 0 & 0 & 0\cr
\langle yz\downarrow| & i\lambda/2 & 0  &  -\lambda/2 & 0 & 0 & 0 \cr
\langle xy\downarrow| & -i\lambda/2 &  -\lambda/2   & -\epsilon & 0 & 0 & 0\cr
\langle xz\downarrow| &0 & 0 & 0 & 0 & i\lambda/2 & i\lambda/2\cr
\langle yz\uparrow| & 0 & 0 & 0 &-i\lambda/2 & 0 & \lambda/2\cr
\langle xy\uparrow|&0 & 0 & 0 &-i\lambda/2 & \lambda/2 &-\epsilon}.
\label{HSOC}
\end{equation}
\end{widetext}

 \begin{table*}[t]
\caption{   \label{Table-Psi}
Energies $E$ and wave functions of the Ir $d$ atomic states in the presence of spin-orbit coupling $\lambda \vec L \cdot \vec S$ and cubic 
($\Delta_{cf}$)  as well as tetragonal crystal fields ($\varepsilon , \delta$ assumed to be $\ll \lambda$ and treated perturbatively, keeping only the linear terms in $\epsilon/ \lambda$).
 The standard $|J_{\rm eff} , m \rangle$ labels 
  for the t$_{2g}$ states are also indicated along with the
 expectation values $\langle L_z \rangle$ and $\langle 2 S_z \rangle$, if the net spin is along $\hat z$, and 
 $\langle L_x \rangle$ and $\langle 2 S_x \rangle$, if the net spin is along $\hat x$.
 The quantity $\xi \equiv \epsilon/ \lambda$ is  the ratio of the tetragonal field $\epsilon$ to the spin-orbit coupling constant $\lambda$.
  }
\begin{ruledtabular}
\begin{tabular}{@{}  r | r | l |  r  |  l |   r | r |  r  |r}
\multicolumn{3}{c|}{Cubic Field ($O_h$)} &   \multicolumn{4}{c }{Tetragonal field ($D_{4h}$)}  \\
\hline
orbital & E &$ |J_{\rm eff}, m \rangle $&  E & wave functions & \multicolumn{2}{c| }{net spin along $\hat{z}$} & \multicolumn{2}{c}{net spin along $\hat{x}$}   \\\cline{6-9}

 & & & & & $\langle 2S_z \rangle$  &$\langle L_z \rangle $&$\langle 2S_x \rangle$  & $\langle L_x \rangle $  \\ \cline{1-9}
 
$e_g$ &$\Delta_{cf} $ & & $\Delta_{cf}+\delta $& $ 3 z^2 - 1\uparrow, 3 z^2 - 1\downarrow$   &$+1, -1$ &0 &$+1, -1$ &0  \\
& &  & $\Delta_{cf}$&     $x^2-y^2\uparrow, x^2-y^2\downarrow$   & $+1, -1$  &0 &$+1, -1$ &0  \\
\hline
 &     $\lambda$ &$ |\frac{1}{2}, \frac{1}{2} \rangle $ 
& $\lambda -\frac{\varepsilon}{3} $  
&$     [ (1+\frac{2\xi}{9})(yz\downarrow + i xz \downarrow)  $& $-\frac{1}{3}-\frac{16\xi}{27}$&$-\frac{2}{3}-\frac{8\xi}{27}$ &$\frac{1}{3}-\frac{8\xi}{27}$
 &$\frac{2}{3}-\frac{4\xi}{27}$ \\

&  &  &  &$+(1-\frac{4\xi}{9})xy\uparrow] / \sqrt 3$& & &\\

&  &$ |\frac{1}{2}, -\frac{1}{2}  \rangle $&  & 
$[ (1+\frac{2\xi}{9})(-yz\uparrow + i xz \uparrow)  $ &$\frac{1}{3}+\frac{16\xi}{27}$&$\frac{2}{3}+\frac{8\xi}{27}$ & $-\frac{1}{3}+\frac{8\xi}{27}$&$-\frac{2}{3}+\frac{4\xi}{27}$\\
 
& &  &  &
$+(1-\frac{4\xi}{9})xy\downarrow] / \sqrt 3$& & &\\\cline{2-9}

$t_{2g}$  &  $-\frac{\lambda}{2}$ & $ |\frac{3}{2}, \frac{1}{2} \rangle $ &
$-\frac{\lambda}{2} - \frac{2\varepsilon}{3}$
 &$[(1-\frac{4\xi}{9})(-yz\downarrow-ixz\downarrow)$ & $\frac{1}{3}+\frac{16\xi}{27}$&$-\frac{1}{3}+\frac{8\xi}{27}$ &$\frac{2}{3}+\frac{8\xi}{27}$&$-\frac{2}{3}+\frac{4\xi}{27}$\\
 & &  &    & 
$+2(1+\frac{2\xi}{9})xy\uparrow]/\sqrt 6$ & & &\\

 &  & $ |\frac{3}{2}, -\frac{1}{2} \rangle $&  &  
 $[(1-\frac{4\xi}{9})(yz\uparrow-ixz\uparrow)$&$-\frac{1}{3}-\frac{16\xi}{27}$&$\frac{1}{3}-\frac{8\xi}{27}$ &$-\frac{2}{3}-\frac{8\xi}{27}$&$\frac{2}{3}-\frac{4\xi}{27}$\\

&& &  & $+2(1+\frac{2\xi}{9})xy\downarrow]/\sqrt 6$&& &\\\cline{3-9}

&  & $ |\frac{3}{2}, \frac{3}{2}\rangle $ &   $\ -\frac{\lambda}{2}$ &  $(yz\uparrow +ixz\uparrow) / \sqrt 2  $  &$1$  &$-1$  &0 &0\\
&  &$ |\frac{3}{2}, -\frac{3}{2} \rangle $ &   &  $(yz\downarrow - ixz\downarrow)/ \sqrt 2$ & $-1$  &$1$ &0 &0\\
\end{tabular}
\end{ruledtabular}
\end{table*}

By diagonalizing the Hamiltonian and keeping the tetragonal field to the lowest order, we obtain the eigenvalues and eigenfuctions, which are summarized in Table \ref{Table-Psi} including some relevant expectation values.
 Note that the wave functions $| J_{\rm eff} , m \rangle$  listed in Table \ref{Table-Psi}   are not necessarily eigenstates of $J^2$ and $J_z$, because they were {\it not} obtained by diagonalizing the  full $10 \times 10$ $\lambda \vec L \cdot \vec S$ matrix, but rather just the $6 \times 6$ matrix in the $t_{2g}$ sector.

The basic electronic structure of the iridates is determined by the Ir$^{4+}$ ions with the $5d^5$ configuration
and a large crystal field due to the oxygen octahedra. The  crystal field combined with the SOC results in the spin-orbital entangled states as summarized in Table \ref{Table-Psi}.
The  cubic  crystal field splits the $5d$ states into  $e_g$ plus $t_{2g}$ states. With the SOC included,  
the six-fold degenerate $t_{2g}$ states (including spin) split into a two-fold 
$J_{\rm eff} = 1/2   $ (higher energy)
and a four-fold   $J_{\rm eff} = 3/2 \  $  (lower energy)
 state.  With the $5d^5$ configuration, four electrons fill the $J_{\rm eff} = 3/2$ state, while the remaining electron
 occupies the lower Hubbard band separated from the upper Hubbard band by the Coulomb interaction $U$ thus resulting in a SOC-induced Mott insulator. 
  The  tetragonal  field  has been  modeled in Table \ref{Table-Psi} by  
asymmetric on-site energies for the Ir d orbitals, viz., 
 $\varepsilon =  \varepsilon (xz/yz) - \varepsilon (xy)    $  and  $\delta = \varepsilon (3 z^2 - 1) - \varepsilon (x^2-y^2) $.
 The $d^5$ configuration is therefore equivalent to a single hole $t_{2g}^1$ with the $ |1/2, 1/2 \rangle$ configuration.

 In Table \ref{Table-Psi}, we have also listed the  angular momentum expectation values,
 which will be useful for the interpretation of the computed magnetic moments 
 discussed in Section \ref{magnetic moment}. 
 The Table lists
   $ \langle L_z \rangle$ and $ \langle 2S_z \rangle$ for the spin state along $\hat z$;
    It is straightforward to show that  $\langle L_x  \rangle = \langle L_y \rangle = \langle S_x  \rangle = \langle S_y \rangle = 0$ in this case.  
Also listed are the expectation values 
 $ \langle L_x \rangle$ and $ \langle 2S_x \rangle$ 
  for the spin state along $\hat x$ direction (which happens to be the case for the ground state of SIO); 
Again, $\langle L_y  \rangle = \langle L_z \rangle = \langle S_y  \rangle = \langle S_z \rangle = 0$ for this case. 
One point to note here is that the tetragonal field $\xi$  mixes up the $J_{\rm eff} = 1/2$ and 3/2 wave functions. 
 For spin magnetization along $\hat z$, the expectation values of the magnetic moments are 
\begin{equation}
\mu_l^z =  \mu_B \langle \psi_{1/2}   |L_z |\psi_{1/2} \rangle =  2/3+8\xi/27,
\end{equation}
\begin{equation}
\mu_s^z= \mu_B  \langle \psi_{1/2}   |2S_z |\psi_{1/2} \rangle=1/3+16\xi/27,
\end{equation}
where $|\psi_{1/2} \rangle = | 1/2, 1/2 \rangle$ , the Bohr magneton $\mu_B = 1$, and other components of the magnetic moments are zero,
leading to the ratio of the magnetic moments to be
\begin{equation}
 \mu_l^z  /   \mu_s^z   =  2-8\xi /3 .
\end{equation}

For Sr$_2$IrO$_4$, the magnetization is along  $\hat x$. 
With the spin wave function along $\hat x$, $| \bar \psi_{1/2} \rangle = 2^{-1/2}  (| 1/2, 1/2 \rangle + | 1/2, -1/2 \rangle )$, we then have
\begin{equation}
\mu_l^{x}  =  \mu_B \langle \bar \psi_{1/2}   |L_x| \bar \psi_{1/2}  \rangle= 2/3-4\xi/27,
\end{equation}
\begin{equation}
\mu_s^{x}      = \mu_B  \langle \bar \psi_{1/2}   |2S_x| \bar \psi_{1/2}  \rangle = 1/3-8\xi/27,
\end{equation}
and the ratio
\begin{equation}
 \mu_l^x /   \mu_s^x   =  2+ 4\xi /3 .
\end{equation}


\end{document}